\documentclass[traditabstract]{aa}
\usepackage{graphicx}
\usepackage{txfonts}
\usepackage{color}
\usepackage{dblfloatfix}

\newcommand{\secp}{\mbox{\rlap{.}$''$}} 
\newcommand{\rasecp}{\mbox{\rlap{.}$^{\rm s}$}} 
\newcommand{\tablenotea}[1]{\parbox{17.5cm}{\indent \footnotesize{#1}}}

\newcommand{\nature}{Nature}

\newcommand{\jms}{J. Mol. Spectr.}

\newcommand{\jpcl}{J. Phys. Chem. Lett.}
\newcommand{\jpca}{J. Phys. Chem. A}
\newcommand{\cpl}{Chem. Phys. Lett.}

\newcommand{\pccp}{Phys. Chem. Chem. Phys.}

\newcommand{\fdis}{Faraday Discuss.}

\newcommand{\jpc}{J. Phys. Chem.}

\newcommand{\pnas}{PNAS}
\newcommand{\psp}{Planet. Space Sci.}
\newcommand{\jesrp}{J. Electron Spectrosc. Relat. Phenom.}
\newcommand{\jppa}{J. Photochem. Photobiol. A}
\newcommand{\rgsp}{Rev. Geophys. Space Phys.}
\newcommand{\molphys}{Mol. Phys.}
\newcommand{\pss}{Planet. Space Sci.}
\newcommand{\hca}{Helvetica Chim. Acta}
\newcommand{\cjp}{Can. J. Phys.}
\newcommand{\scaa}{Spectrochim. Acta Part A}
\newcommand{\bcsj}{Bull. Chem. Soc. Japan}

\begin{document}

\title{The growth of carbon chains in IRC\,+10216 mapped with ALMA\thanks{Based on observations carried out with ALMA and the IRAM 30m Telescope. ALMA is a partnership of ESO (representing its member states), NSF (USA) and NINS (Japan), together with NRC (Canada) and NSC and ASIAA (Taiwan), in cooperation with the Republic of Chile. The Joint ALMA Observatory is operated by ESO, AUI/NRAO and NAOJ. IRAM is supported by INSU/CNRS (France), MPG (Germany) and IGN (Spain). This paper makes use of the following ALMA data: ADS/JAO.ALMA\#2013.1.00432.S.}}

\titlerunning{The growth of carbon chains in IRC\,+10216}
\authorrunning{Ag\'undez et al.}

\author{
M.~Ag\'undez\inst{1}, J.~Cernicharo\inst{1}, G.~Quintana-Lacaci\inst{1}, A.~Castro-Carrizo\inst{2}, L.~Velilla Prieto\inst{1}, N.~Marcelino\inst{1}, M.~Gu\'elin\inst{2}, C. Joblin\inst{3,4}, J. A. Mart\'in-Gago\inst{1}, C.~A.~Gottlieb\inst{5}, N.~A.~Patel\inst{5}, and M. C.~McCarthy\inst{5,6}
}

\institute{
Instituto de Ciencia de Materiales de Madrid, CSIC, C/ Sor Juana In\'es de la Cruz 3, 28049 Cantoblanco, Spain \and
Institut de Radioastronomie Millim\'etrique, 300 rue de la Piscine, 38406 St. Martin d'H\'eres, France \and
Universit\'e de Toulouse, UPS-OMS, IRAP, 31000 Toulouse, France \and
CNRS, IRAP, 9 Av. Colonel Roche, BP 44346, 31028 Toulouse Cedex 4, France \and
Harvard-Smithsonian Center for Astrophysics, 60 Garden Street, Cambridge, MA 02138, USA \and
School of Engineering and Applied Sciences, Harvard University, Cambridge, MA 02138, USA
}

\date{Received; accepted}


\abstract
{Linear carbon chains are common in various types of astronomical molecular sources. Possible formation mechanisms involve both bottom-up and top-down routes. We have carried out a combined observational and modeling study of the formation of carbon chains in the C-star envelope IRC\,+10216, where the polymerization of acetylene and hydrogen cyanide induced by ultraviolet photons can drive the formation of linear carbon chains of increasing length. We have used ALMA to map the emission of $\lambda$~3~mm rotational lines of the hydrocarbon radicals C$_2$H, C$_4$H, and C$_6$H, and the CN-containing species CN, C$_3$N, HC$_3$N, and HC$_5$N with an angular resolution of $\sim$1$''$. The spatial distribution of all these species is a hollow, 5-10$''$ wide, spherical shell located at a radius of 10-20$''$ from the star, with no appreciable emission close to the star. Our observations resolve the broad shell of carbon chains into thinner sub-shells which are 1-2$''$ wide and not fully concentric, indicating that the mass loss process has been discontinuous and not fully isotropic. The radial distributions of the species mapped reveal subtle differences: while the hydrocarbon radicals have very similar radial distributions, the CN-containing species show more diverse distributions, with HC$_3$N appearing earlier in the expansion and the radical CN extending later than the rest of the species. The observed morphology can be rationalized by a chemical model in which the growth of polyynes is mainly produced by rapid gas-phase chemical reactions of C$_2$H and C$_4$H radicals with unsaturated hydrocarbons, while cyanopolyynes are mainly formed from polyynes in gas-phase reactions with CN and C$_3$N radicals.}
{}
{}
{}
{}


\maketitle

\section{Introduction}

Linear acetylenic carbon chains, with a polyyne ($-$C$\equiv$C)$_n$$-$H or cyanopolyyne ($-$C$\equiv$C)$_n$$-$C$\equiv$N backbone, have long been observed in a wide range of astronomical environments, such as cold molecular clouds, star-forming regions, photon-dominated regions, and the ejecta of evolved stars (\cite{mor1976} 1976; \cite{win1978} 1978; \cite{gue1978} 1978; \cite{bro1978} 1978; \cite{buj1981} 1981; \cite{cer1996} 1996; \cite{par2005} 2005; \cite{gup2009} 2009; see \cite{loo2016} 2016 for a recent disproof of HC$_{11}$N detection). The chemical synthesis of carbon chains in these environments has been a matter of debate over the years, with hypotheses based on top-down mechanisms involving very small carbonaceous grains or aromatic polycyclic hydrocarbons as precursors (\cite{kro1987} 1987; \cite{pet2005} 2005), and bottom-up chemical schemes driven by cosmic rays or ultraviolet photons (\cite{her1989} 1989; \cite{fuk1998} 1998; \cite{cer2004} 2004).

Carbon chains are particularly conspicuous in circumstellar envelopes around carbon-rich Asymptotic Giant Branch (AGB) stars, which are also the main factories of carbonaceous dust grains in the Galaxy. Circumstellar envelopes provide an ideal laboratory in which to study the formation of carbon chains because of their relative simplicity in geometry, velocity field, and composition of the precursor material. Because chemical processes take place as matter is ejected from the star through a nearly isotropic outflow, the sequential formation of molecules can be tracked as a function of the radial position in the envelope. The central AGB star is not expected to emit at ultraviolet wavelengths and thus the main source of ultraviolet photons is the ambient radiation field from nearby stars. Chemical models point to a bottom-up scenario in which the formation of carbon chains of increasing length occurs because of ultraviolet-driven polymerization of C$_2$H$_2$ and HCN. These small precursors are injected into the expanding wind from the inner circumstellar regions and are photodissociated in the outer circumstellar layers, where the external ambient ultraviolet field is no longer blocked by the circumstellar dust and gas. In these outer regions, the production of the radicals C$_2$H and CN drives the growth of acetylenic carbon chains (\cite{gla1986} 1986; \cite{nej1987} 1987; \cite{che1993a} 1993; \cite{che1993b} 1993; \cite{mil1994} 1994; \cite{mil2000} 2000; \cite{agu2009} 2009; \cite{cor2009} 2009; \cite{li2014} 2014).

Owing to its close proximity and brightness, most observational studies of circumstellar envelopes have focused on the prototypical carbon star envelope IRC\,+10216. In this source, brightness distributions of rotational lines of the radicals C$_2$H and CN, and polyyne and cyanopolyyne chains such as C$_4$H, C$_3$N, HC$_3$N, and HC$_5$N have been mapped using single dish radiotelescopes such as IRAM 30m (\cite{aud1994} 1994) and radio interferometers such as BIMA (\cite{bie1993} 1993; \cite{day1995} 1995), IRAM Plateau de Bure (\cite{gue1993} 1993, 1997, 1999; \cite{luc1995} 1995), VLA (\cite{din2008} 2008), and SMA (\cite{pat2011} 2011; \cite{coo2015} 2015). Despite their moderate sensitivity and angular resolution (ranging from 2$''$ to 11$''$), these observations were able to constrain the presence of C$_2$H, CN, and the aforementioned carbon chains to an outer hollow shell of a few arcseconds of width located at a distance between 10$''$ and 20$''$ from the star. This finding implies that whatever the mechanism of formation of carbon chains in IRC\,+10216, it is only activated in a very specific region of the outer envelope. However, there are still many open questions such as whether all carbon chains occur precisely in the same circumstellar region or whether there is some degree of radial stratification depending on the carbon chain length. A deeper look into the exact chemical mechanisms behind the formation of carbon chains requires observations with a significantly higher sensitivity and angular resolution.

In this article, we present sensitive ALMA observations of IRC\,+10216 in several rotational lines of carbon chains across the $\lambda$~3~mm band. The observations have an angular resolution of $\sim$1$''$, which allows us to resolve the detailed structure and to precisely locate the circumstellar regions where the emission of each carbon chain arises. A chemical model has been developed to interpret the observations and to explain the carbon chain chemistry that is taking place in the outer envelope of this archetypal carbon star.

\section{Observations} \label{sec:observations}

A $\lambda~3$~mm spectral survey of IRC\,+10216 covering the frequency range 84.0-115.5 GHz was carried out with ALMA band\,3 during Cycle\,2. Observations were obtained with compact and extended array configurations. The field of view (primary beam) of the 12m ALMA antennas ranges from $\sim$69$''$ at 84 GHz to $\sim$50$''$ at 115.5 GHz. Additional observations were performed with the IRAM 30m telescope to recover the flux filtered out by the interferometer. Observations were centered on the position of the star, with coordinates J2000.0 R.A.=09$^{\rm h}$47$^{\rm m}$57\rasecp446 and Dec.=13$^{\circ}$16$'$43\secp86, according to the position of the $\lambda$~1~mm continuum emission peak (\cite{cer2013} 2013). A thorough description of the spectral survey will be presented elsewhere (Cernicharo et al., in preparation). Here, we focus on the large-scale emission of carbon chains. For the emission lines studied here, data from the ALMA compact and extended configurations were merged, after continuum subtraction, with the short-spacings data obtained with IRAM 30m, providing an angular resolution of $\sim$1$''$.

\begin{figure*}
\begin{minipage}{\columnwidth}
\begin{flushleft} \vspace{-5.70cm}
\includegraphics[angle=0,width=0.92\columnwidth]{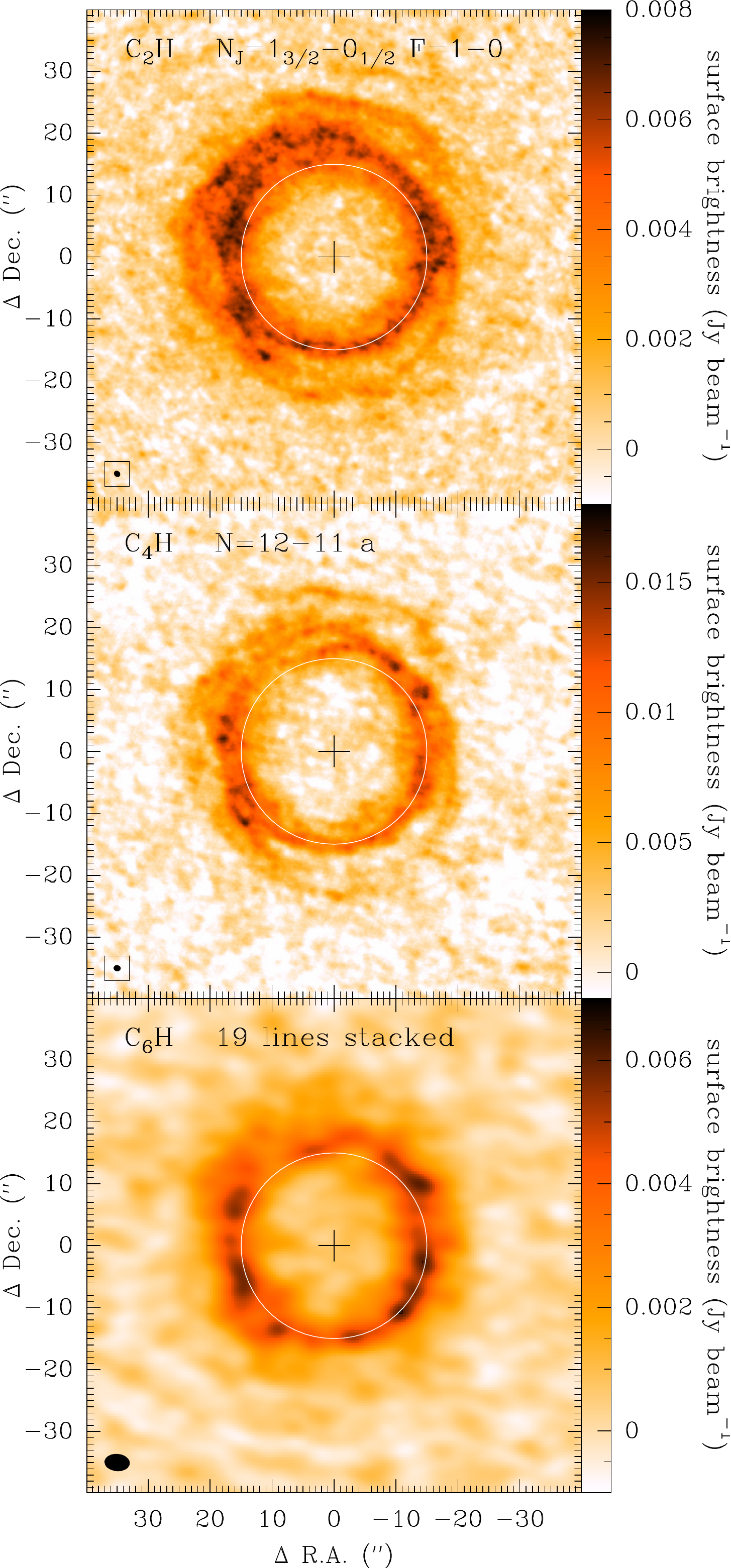}
\end{flushleft}
\end{minipage}
\begin{minipage}{\columnwidth}
\begin{flushright}
\includegraphics[angle=0,width=0.92\columnwidth]{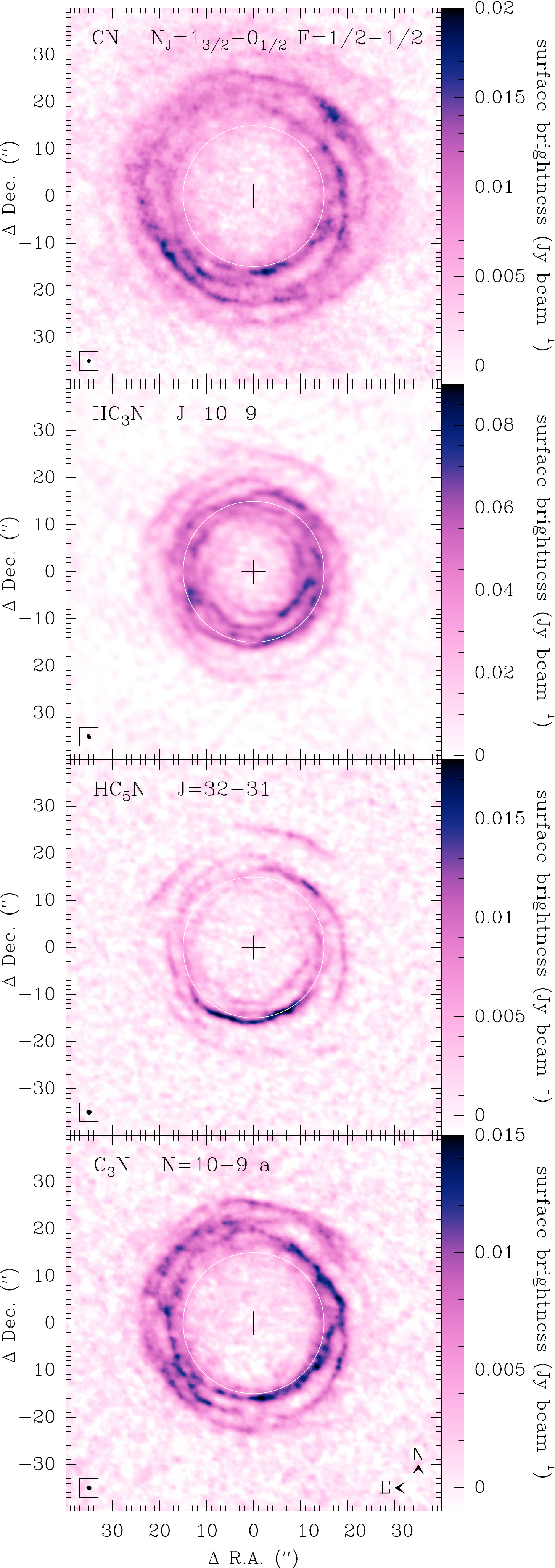}
\end{flushright}
\end{minipage}
\caption{Brightness distributions of $\lambda$ 3~mm rotational transitions of the polyyne-like radicals C$_2$H, C$_4$H, and C$_6$H, and the cyanopolyyne-like species CN, HC$_3$N, HC$_5$N, and C$_3$N, averaged over a velocity range of width 3 km s$^{-1}$ centered at $V_{\rm LSR}=V_{\rm sys}$. Details on the maps can be found in Section~\ref{sec:observations}. The maps are centered on the star, indicated by a cross, and the synthesized beam ($\sim$1$''$ in all maps but that of C$_6$H, where it is $\sim$3$''$) is shown in the bottom left corner of each panel. A white circle of radius 15$''$ centered on the star is drawn to facilitate comparison among maps.} \label{fig:map_all}
\end{figure*}

There are several emission lines from carbon chains in the $\lambda~3$~mm ALMA spectrum of IRC\,+10216. In this article we focus on linear carbon chains of the families of polyynes and cyanopolyynes. With respect to polyynes, we concentrate on the radicals of increasing length C$_2$H, C$_4$H, and C$_6$H, which belong to a chemical series resulting from the addition of C$_2$ subunits. These radicals are chemical descendants of acetylene and the symmetric polyynes C$_4$H$_2$ and C$_6$H$_2$, molecules that lack a permanent electric dipole moment and thus cannot be observed at these wavelengths. For cyanopolyynes, we study the emission of the radical CN, the cyanopolyynes HC$_3$N and HC$_5$N, and the related radical C$_3$N. We concentrated on these seven species because they display intense emission lines in the $\lambda~3$~mm spectrum of IRC\,+10216 and share a close chemical relationship as precursors or members of the family of acetylenic carbon chains.

An overview of the spatial distribution of C$_2$H, C$_4$H, and C$_6$H (left side) and CN, HC$_3$N, HC$_5$N, and C$_3$N (right side) is shown in Fig.~\ref{fig:map_all}, where we plot the brightness distribution of a representative line of each species, averaged over a velocity range of 3~km s$^{-1}$ centered at $V_{\rm LSR}=V_{\rm sys}$ (where $V_{\rm sys} = -26.5$ km s$^{-1}$ for IRC\,+10216; \cite{cer2000} 2000). The choice of this velocity range is very convenient for spherically expanding envelopes such as IRC\,+10216 because the maps show the brightness distribution in a plane perpendicular to the line of sight intersecting the star, allowing us to identify the radial dependence of the emission.

\textbf{C$_2$H and CN.} For the small radicals C$_2$H and CN, only the $N=1-0$ rotational transition of each falls within the frequency range of the survey. This transition in turn consists of two spin-rotation fine structure components, which in turn are further split into several hyperfine components due to the nuclear spin of the hydrogen (C$_2$H) or nitrogen (CN) nucleus. Some of the hyperfine components which are blended in the spectrum of IRC\,+10216 are closer in frequency than half the linewidth, which is given by the terminal expansion velocity of the envelope (14.5 km s$^{-1}$; \cite{cer2000} 2000), implying that the emission at $V_{\rm LSR}=V_{\rm sys}$ for one component is contaminated by blue- or red-shifted emission from the overlapping component. These components are therefore not adequate to trace the spatial distribution of C$_2$H or CN in the plane of the sky. In the top left panel of Fig.~\ref{fig:map_all} we show the spatial distribution of C$_2$H through a map of the $J=3/2-1/2$ $F=1-0$ component, lying at 87328.625 MHz, which is the most intense free of blending component. In this map the synthesized beam is $1\secp07\times0\secp92$ and the rms per 3 km s$^{-1}$ channel is 0.4 mJy beam$^{-1}$. In the case of CN, there are various intense non-overlapping components that show similar brightness distributions. The spatial distribution of CN is illustrated in the top right panel of Fig.~\ref{fig:map_all} through the brightness distribution of the $J=3/2-1/2$ $F=1/2-1/2$ component, whose rest frequency is 113499.644 MHz. For this map, the synthesized beam is $0\secp85\times0\secp74$ and the rms per 3 km s$^{-1}$ channel is 1.0 mJy beam$^{-1}$.

\textbf{C$_4$H and C$_3$N.} The radicals C$_4$H and C$_3$N have four and three, respectively, rotational transitions (each consisting of a spin-rotation doublet) within the frequency range 84.0-115.5 GHz. The brightness distributions of the individual lines of the two species are discussed in Appendix~\ref{sec:map_trans}. In Fig.~\ref{fig:map_all} we show the emission map of a selected line of each of these radicals to illustrate their spatial distributions in IRC\,+10216. The maps of C$_4$H and C$_3$N in Fig.~\ref{fig:map_all} have a similar angular resolution, $1\secp14\times0\secp98$ and $1\secp10\times0\secp92$, respectively, and sensitivity, with rms levels per 3 km s$^{-1}$ channel of 1.3 and 0.7 mJy beam$^{-1}$, respectively.

\textbf{HC$_3$N and HC$_5$N.} The cyanopolyynes HC$_3$N and HC$_5$N have several rotational transitions within the frequency range covered, three in the case of HC$_3$N and up to 12 for HC$_5$N. A detailed discussion on the brightness distributions of the individual lines of the two molecules is also presented in Appendix~\ref{sec:map_trans}. In Fig.~\ref{fig:map_all}, the spatial distributions of HC$_3$N is traced by the emission of the $J=10-9$ line in a map with an angular resolution of $1\secp08\times0\secp81$ and a rms per 3 km s$^{-1}$ channel of 0.9 mJy beam$^{-1}$, while the distribution of HC$_5$N is illustrated by the emission of the $J=32-31$ line, with an angular resolution of $1\secp11\times0\secp99$ and a rms per 3 km s$^{-1}$ channel of 0.6 mJy beam$^{-1}$.

\begin{figure*}
\centering
\includegraphics[angle=0,width=0.78\textwidth]{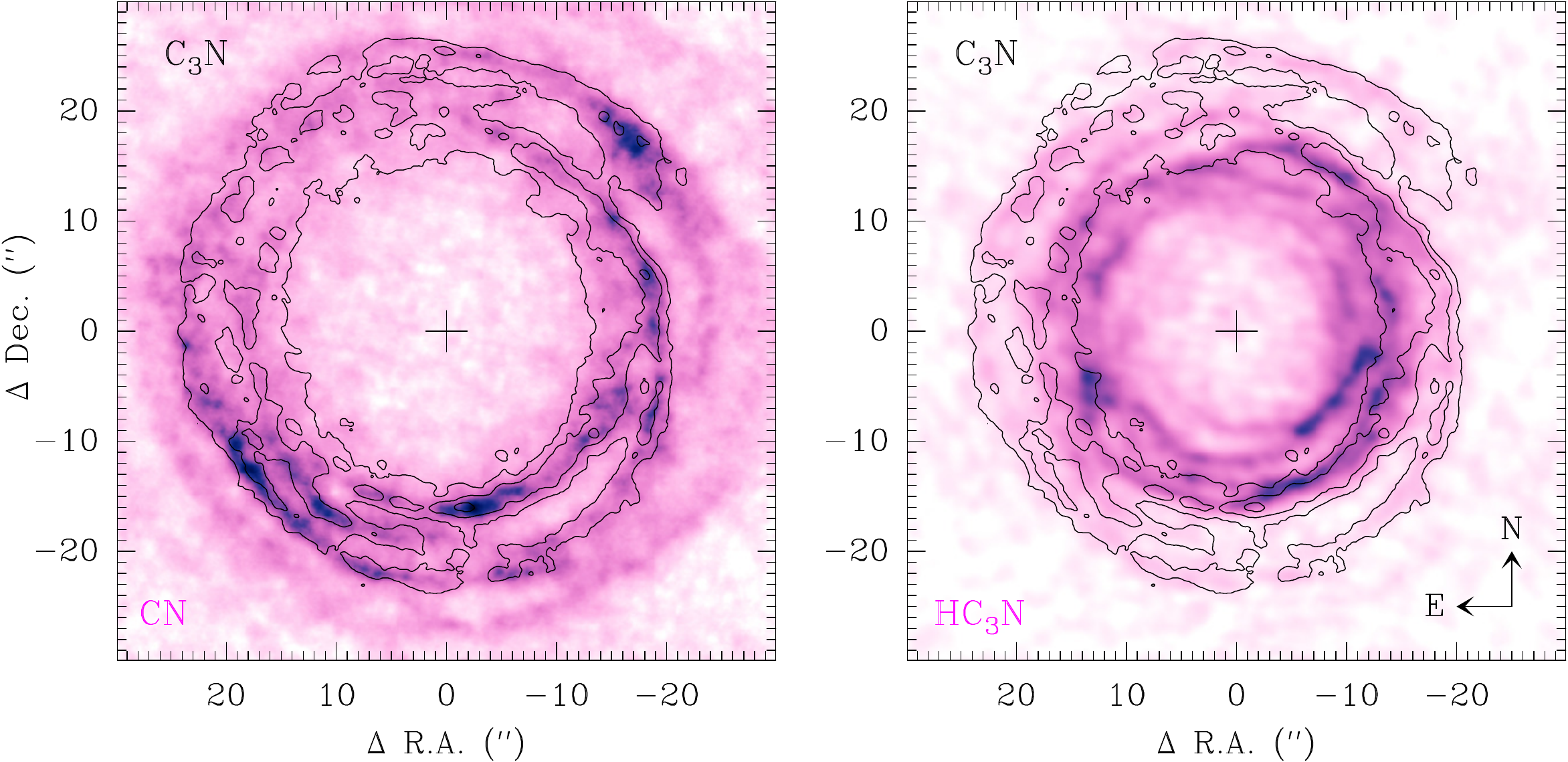}
\caption{Emission distribution of C$_3$N $N=10-9$ (black contours) superimposed on the brightness distribution of CN $N=1-0$ (color map in the left panel) and of HC$_3$N $J=10-9$ (color map in the right panel).} \label{fig:map_superimposed}
\end{figure*}

\textbf{C$_6$H.} The radical hexatriynyl, C$_6$H, has a plethora of rotational transitions in the $\lambda~3$~mm spectrum of IRC\,+10216, consisting of two series of doublets arising from the $^2\Pi_{1/2}$ and $^2\Pi_{3/2}$ spin-orbit fine structure states, the former lying 21.6 K above the latter. Due to the lower sensitivity of the interferometric data of each individual line, we only considered ALMA data obtained with the compact configuration and stacked the maps of a selected list of 19 lines, all free of blending and observed with a sufficient quality, to obtain a more sensitive map of the spatial distribution of C$_6$H (see Fig.~\ref{fig:map_all}). Because we have omitted the ALMA data obtained with the extended configuration, the angular resolution of the C$_6$H map is of the order of $\sim$3$''$, i.e., significantly poorer than in the rest of maps shown in Fig.~\ref{fig:map_all}. The set of C$_6$H lines stacked cover a wide range of upper level energies, 64-86 K for lines from the $^2\Pi_{3/2}$ state and 86-113 K for $^2\Pi_{1/2}$ lines, and thus it is possible that their different excitation requirements result in different brightness distributions. However, we note that on the basis of the behaviour of other molecules such as HC$_5$N (see Appendix~\ref{sec:map_trans}), such differences are unlikely to be larger than the current angular resolution of $\sim$3$''$.

\section{Observed spatial distributions of carbon chains}

A cursory view of the maps shown in Fig.~\ref{fig:map_all} reveals a clear and well-defined ring structure in the seven species studied: the three hydrocarbon radicals C$_2$H, C$_4$H, and C$_6$H, and the four cyanides CN, HC$_3$N, HC$_5$N, and C$_3$N. The maps correspond to the emission at the line center, i.e., at $V_{\rm LSR}=V_{\rm sys}$, and thus indicate how the emission is distributed along a plane perpendicular to the line of sight intersecting the star. The two-dimensional ring structure, together with the emission maps at red- and blue-shifted velocities (not shown in Fig.~\ref{fig:map_all}), indicate that the three-dimensional structure of the seven species consists of a hollow spherical shell with a width of 5-10$''$ located at a distance between 10$''$ and 20$''$ from the star. This distribution was already known for most species studied here from previous interferometric maps, albeit at lower angular resolution (see \cite{gue1997} 1997 and references therein).

The higher angular resolution of the ALMA observations presented here reveals that in fact the broad, 5-10$''$ wide, hollow shell consists of several clumpy thinner shells of width 1-2$''$, which are not fully concentric, as indicated by the arcs which cross each other and by a spiral-like structure seen in the $\sim$1$''$ resolution maps of Fig.~\ref{fig:map_all}. This type of substructure was seen earlier in IRC\,+10216 at larger scales in light scattered by dust grains (\cite{mau1999} 1999; \cite{lea2006} 2006) and in emission from CO (\cite{cer2015} 2015), and has been interpreted in terms of episodic or discontinuous mass loss owing to the presence of a binary companion (\cite{cer2015} 2015). Observations carried out with the VLA at an angular resolution of $\sim$2-3$''$ also showed that the cm-wavelength emission of HC$_3$N and HC$_5$N consists of clumpy shells consistent with a non-isotropic and episodic mass loss process (\cite{din2008} 2008). More recently, high angular resolution ALMA observations of the emission in different molecular lines have also provided hints of departures from the spherical symmetry and spiral-like structures in the inner regions of the envelope (\cite{dec2015} 2015; \cite{vel2015} 2015; \cite{agu2015} 2015; \cite{qui2016} 2016). The $\sim$1$''$ angular resolution maps (see Fig.~\ref{fig:map_all} using white circles as a reference) show that the molecular shell is not exactly centered on the position of the star but shifted by a few arcseconds in the NE direction. It seems as though the shell of carbon chains is compressed from the SW direction, a result that is also observed at larger scales in CO (\cite{cer2015} 2015). It is currently unclear whether this feature is the result of an enhanced ultraviolet illumination from the direction of the Galactic plane (SW) or it is linked to the non isotropic character of the mass loss process. It is noteworthy that, unlike some of the previous observations at lower angular resolution (see, for example, the case of C$_2$H in \cite{gue1997} 1997), the maps in Fig.~\ref{fig:map_all} show no evidence of axial symmetry.

The structure of thin shells and inter-crossing arcs is common to the emission distribution of the different species studied (at least for those mapped with $\sim$1$''$ angular resolution; see Fig~\ref{fig:map_all}). These similarities indicate that the maxima of brightness correspond to locations where there is an enhancement in the total volume density of matter, and not in the fractional abundance of a particular species. Such a finding is consistent with some mechanism that has shaped the envelope by driving the mass loss process out of the strictly uniform isotropic character.

The brightness distributions of different species reveal subtle differences. Even if there is a good match between the arcs and shells seen for different species, when one moves radially from the star some arcs or thin shells that are observed in some species are not seen in others, which provides information about the chemical particularities of each species. This point is illustrated in Fig.~\ref{fig:map_superimposed}, where we superimpose the emission distribution of the radical C$_3$N, shown as contours, on the brightness distribution of the small radical CN and of the cyanopolyyne HC$_3$N. Referring first to the left panel of Fig.~\ref{fig:map_superimposed}, the shells and arcs in C$_3$N are well matched by CN, although it is also true that CN shows shells beyond $\sim$20$''$ which are not seen in C$_3$N. That is, the CN radical extends farther from the star than the radical C$_3$N. Referring now to the right panel in Fig.~\ref{fig:map_superimposed}, we note again that the shells and arcs in C$_3$N are well matched by similar structures in HC$_3$N. Also note that the well-defined outer arcs indicated by the lowest contour in the NW and SW directions in C$_3$N are well matched by tenuous, although appreciable, features in HC$_3$N. In this case, HC$_3$N shows shells inward of $\sim$14$''$ that are not seen in its radical C$_3$N, which leads us to conclude that, in the framework of an expanding envelope, the cyanopolyyne HC$_3$N appears in regions where the radical C$_3$N has not yet been formed.

\begin{figure}
\centering
\includegraphics[angle=0,width=\columnwidth]{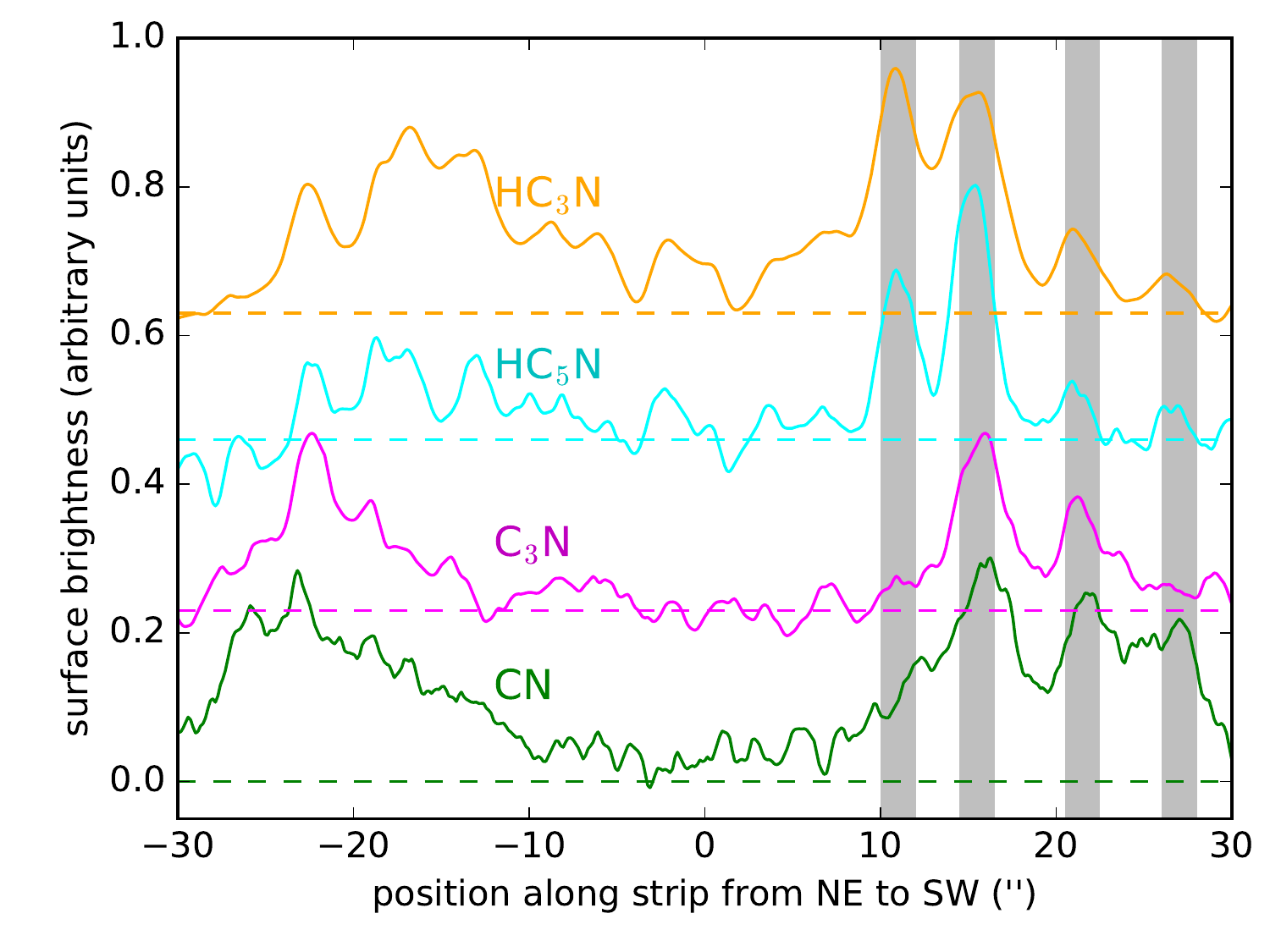}
\caption{Brightness distributions of the cyanides HC$_3$N, HC$_5$N, C$_3$N, and CN along a strip intersecting the star at PA = 45$^\circ$. Curves correspond to the maps shown in Fig.~\ref{fig:map_all} and have been arbitrarily scaled and shifted in the $y$-axis for display. Horizontal dashed lines indicate the zero of intensity. Grey vertical shadows indicate the positions of brightness maxima in the SW direction from the star, at 11$''$, $15\secp5$, $21\secp5$, and 27$''$.} \label{fig:strip}
\end{figure}

Fig.~\ref{fig:strip} shows in a different way how emission distributions of different species match each other. In this plot we have extracted the surface brightness along a strip intersecting the star at PA = 45$^\circ$ from the maps of the cyanides CN, C$_3$N, HC$_3$N, and HC$_5$N shown in Fig.~\ref{fig:map_all}. There are several aspects that are worthy of comment. First, the brightness distributions along the strip selected are not symmetrical with respect to the position of the star, indicating that the circumstellar shells are not strictly concentric, but instead consist of intersecting arcs and a spiral-like structure. Second, if we focus on the SW direction (positive values along $x$-axis), there are several shells of width $\sim$2$''$ at approximately +11$''$, $+15\secp5$, $+21\secp5$, and +27$''$ from the star. These shells are however not traced by the four species. The shells at +$15\secp5$ and +$21\secp5$ are seen in the four species. The shell at +11$''$ is traced by the cyanopolyynes HC$_3$N and HC$_5$N, but not by the radicals C$_3$N and CN, and the farthest one at +27$''$ is mostly seen in CN and to a lesser extent in the cyanopolyynes. Therefore, among the four cyanides, the cyanopolyynes appear at shorter radii than their radicals, while the radical CN is the one that extends farther in the envelope.

\begin{figure}
\centering
\includegraphics[angle=0,width=\columnwidth]{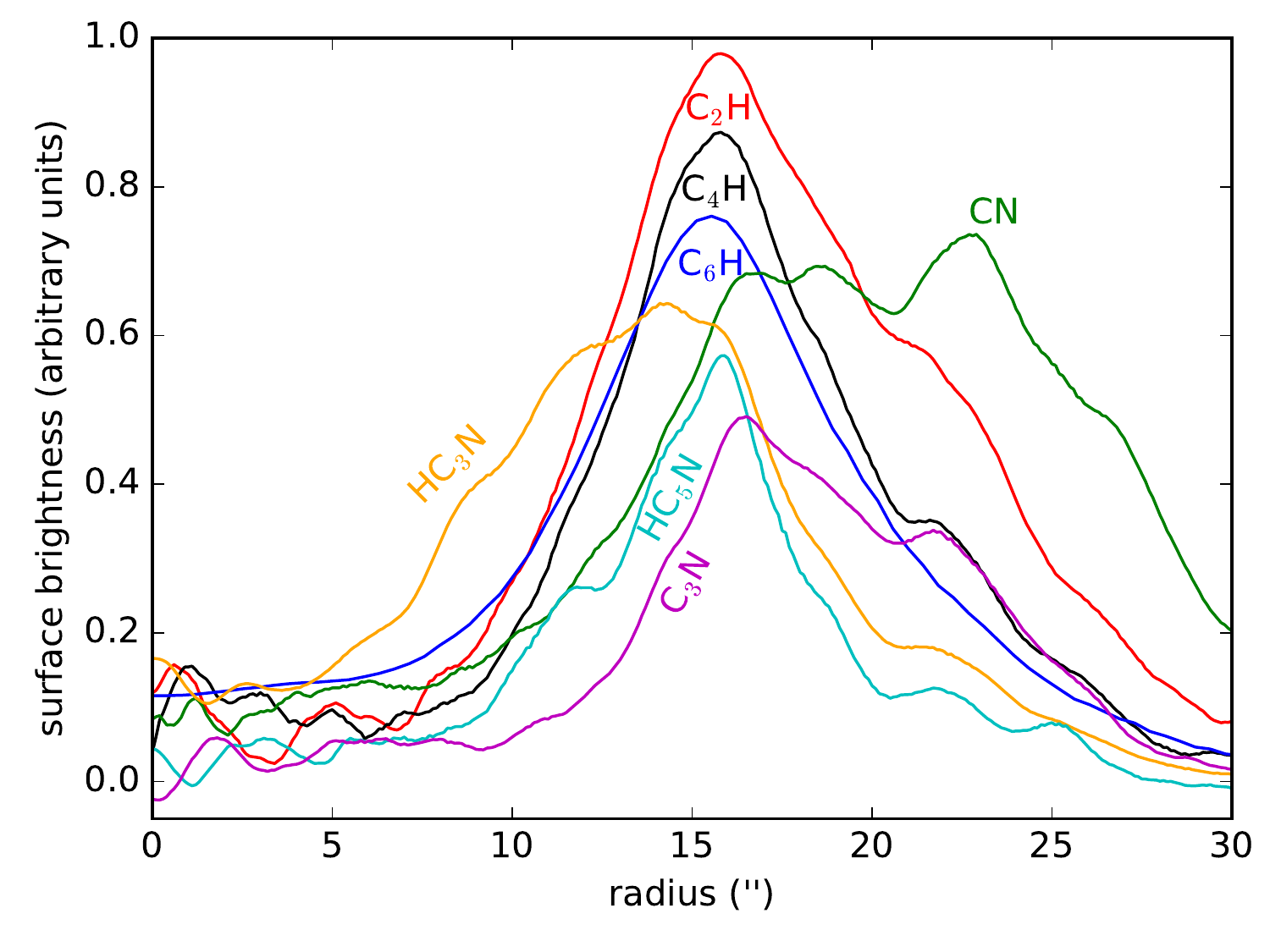}
\caption{Radial brightness distributions of the hydrocarbon radicals C$_2$H, C$_4$H, and C$_6$H, and of the cyanides CN, HC$_3$N, HC$_5$N, and C$_3$N, calculated by azimuthal average of the brightness distributions shown in Fig.~\ref{fig:map_all}. Curves have been arbitrarily scaled in the $y$-axis for display.} \label{fig:azimuth_average}
\end{figure}

Information on the radial distribution of the species can be obtained by azimuthal average of the brightness distribution (see Fig.~\ref{fig:azimuth_average}). The lack of strictly concentric shells, with arcs that cross each other within a spiral-like arrangement, produces a smoothing effect when azimuthally averaging. That is, the spatial resolution is degraded by averaging shells that are not strictly at the same radial position. Still, the information provided by this plot is of great utility in analyzing whether species share the same radial regions or whether there is some degree of stratification as a function of radius. The hydrocarbon radicals C$_2$H, C$_4$H, and C$_6$H reach their maximum brightness at the same radial position, between $15\secp5$ and 16$''$, although the smallest one, C$_2$H, extends somewhat farther than the heavier ones. The cyanides, however, show much more disparate radial distributions. The distribution of HC$_3$N is the one which is most shifted to inner regions, as it emerges at radii as short as 7$''$ while the rest emerge at radial distances greater than 10$''$. The radial distribution of CN is also markedly different from the rest of the species as it extends out to radial distances beyond 30$''$. As for HC$_5$N, it appears at the same radial distance than CN and disappears with HC$_3$N, while the radical C$_3$N appears somewhat farther than the cyanopolyynes HC$_3$N and HC$_5$N. All this information is of great value in understanding the formation mechanism of each of these species in IRC\,+10216 and will be the subject of further discussion in Section~\ref{sec:model}.

\section{Model} \label{sec:model}

The ALMA emission maps of carbon chains obtained toward IRC\,+10216 provides an excellent opportunity to determine the main chemical processes responsible for their formation along the circumstellar outflow. To this end we have constructed a simple model of IRC\,+10216 that consists of a spherical envelope of gas and dust which expands around an AGB star with a constant velocity. The parameters of the star and envelope have been taken from \cite{agu2012} (2012), i.e., the adopted mass loss rate is $2\times10^{-5}$ $M_{\odot}$ yr$^{-1}$. For the distance, we use the recent value of 123 pc determined by \cite{gro2012} (2012). First we constructed a chemical model to describe the chemical composition of the gas as it expands, and then we carried out radiative transfer calculation to convert the abundance radial distributions into brightness radial distributions which can be compared with the ALMA observations.

\subsection{Chemical model: formation of carbon chains} \label{sec:chemical_model}

In the chemical model, the gas is assumed to expand isotropically from an initial radius of $2\times10^{14}$ cm. We consider as parent molecules at the initial radius H$_2$, CO, C$_2$H$_2$, CH$_4$, C$_2$H$_4$, H$_2$O, N$_2$, HCN, NH$_3$, CS, H$_2$S, SiS, SiO, SiH$_4$, PH$_3$, and HCP, with abundances from \cite{agu2012} (2012) and a value of $4\times10^{-5}$ relative to H$_2$ for N$_2$, based on chemical equilibrium calculations and a solar elemental abundance for nitrogen. Of special interest for the formation of carbon chains in the outer envelope are the parent molecules C$_2$H$_2$ and HCN, for which the adopted abundances relative to H$_2$ are $8\times10^{-5}$ and $4\times10^{-5}$, respectively (\cite{fon2008} 2008). We assume that the envelope is externally illuminated by the local ultraviolet radiation field of \cite{dra1978} (1978). Under the assumption of a smooth envelope and adopting the $N_{\rm H}$ / $A_V$ ratio of $1.87\times10^{21}$ cm$^{-2}$ mag$^{-1}$ determined for the local interstellar medium by \cite{boh1978} (1978), chemical models of IRC\,+10216 predict that photochemistry occurs at shorter radii than indicated by interferometric observations (see, e.g., discussion in Appendix B of \cite{agu2006} 2006 and Figures 1 and 4 in \cite{cor2009} 2009). For the purpose of reproducing the observed radial location of carbon chains, we have adopted a $N_{\rm H}$ / $A_V$ ratio 1.5 times lower than the canonical interstellar value of \cite{boh1978} (1978), regardless of the implicit assumption on the dust opacity. The adopted cosmic-ray ionization rate of H$_2$ is $1.2\times10^{-17}$ s$^{-1}$, a value typical of the local interstellar medium that is well suited for IRC\,+10216, according to observations of HCO$^+$ (\cite{agu2006} 2006; \cite{pul2011} 2011). We use a large network of gas-phase chemical reactions, whose rate constants have been taken from the literature on gas-phase chemical kinetics and from the UMIST and KIDA databases (\cite{mce2013} 2013; \cite{wak2015} 2015). More details on the reactions important for the chemistry of carbon chains in IRC\,+10216 and on the rate constants are given below.

\begin{figure*}
\centering
\includegraphics[angle=0,width=\columnwidth]{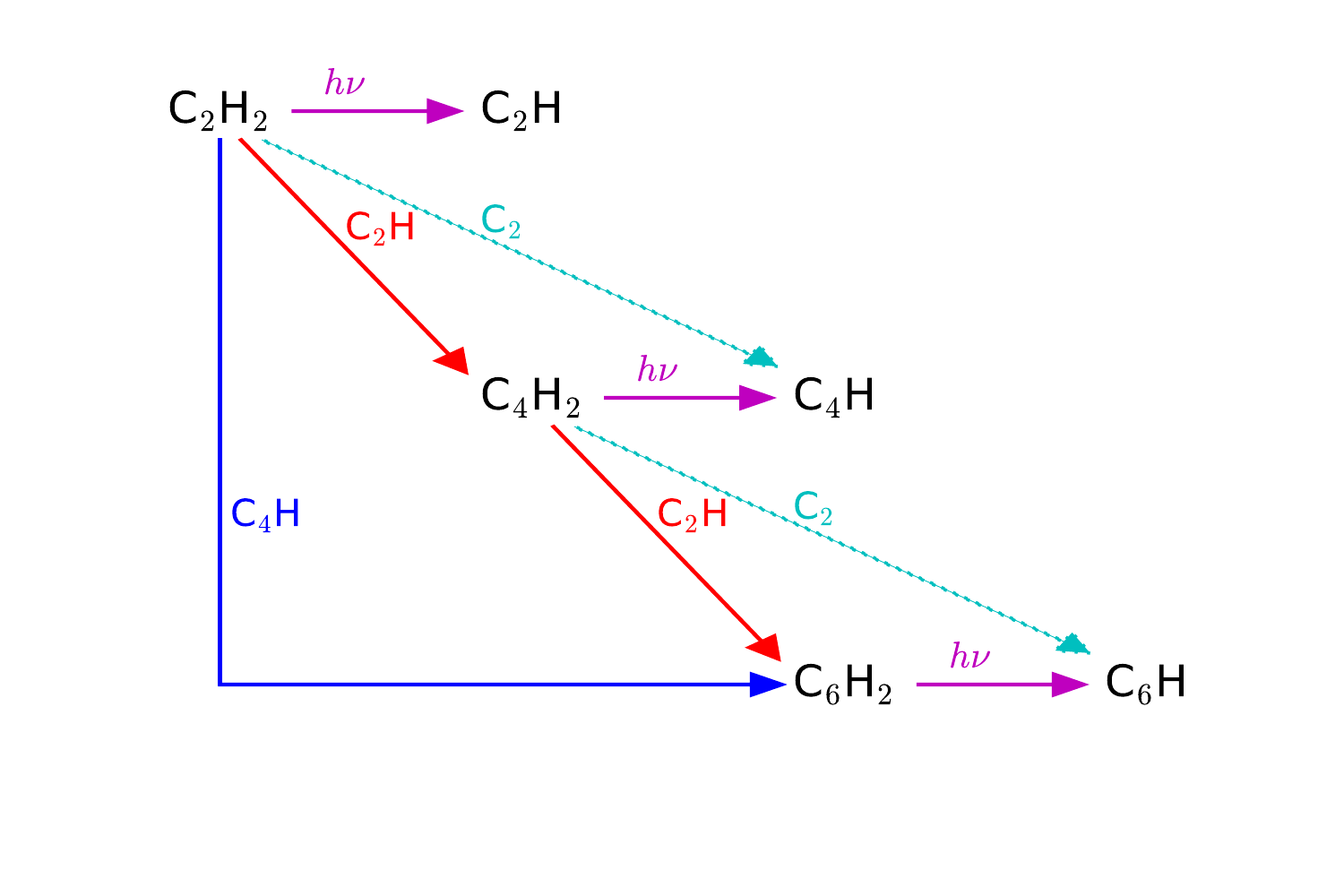}
\includegraphics[angle=0,width=\columnwidth]{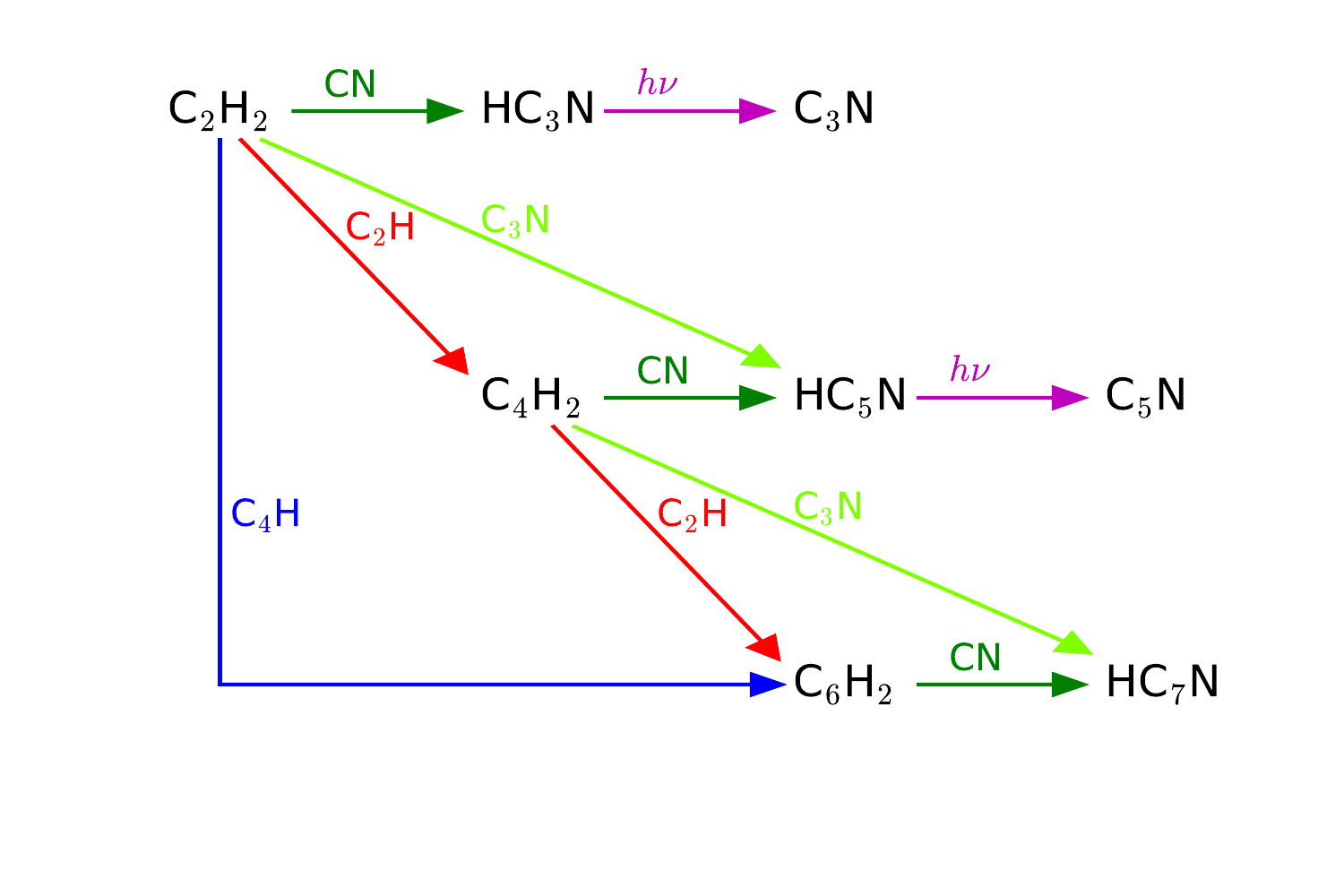}
\includegraphics[angle=0,width=\columnwidth]{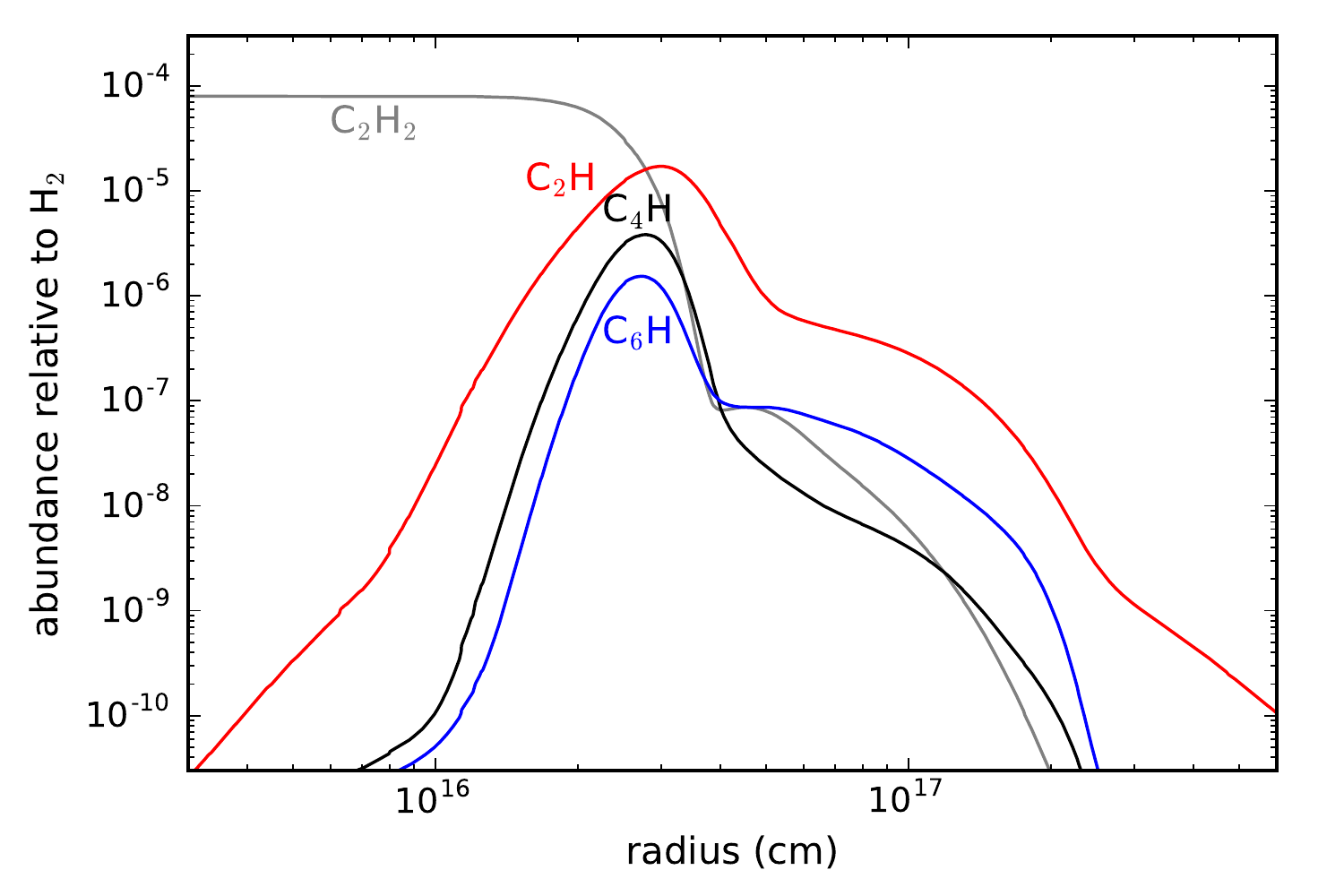}
\includegraphics[angle=0,width=\columnwidth]{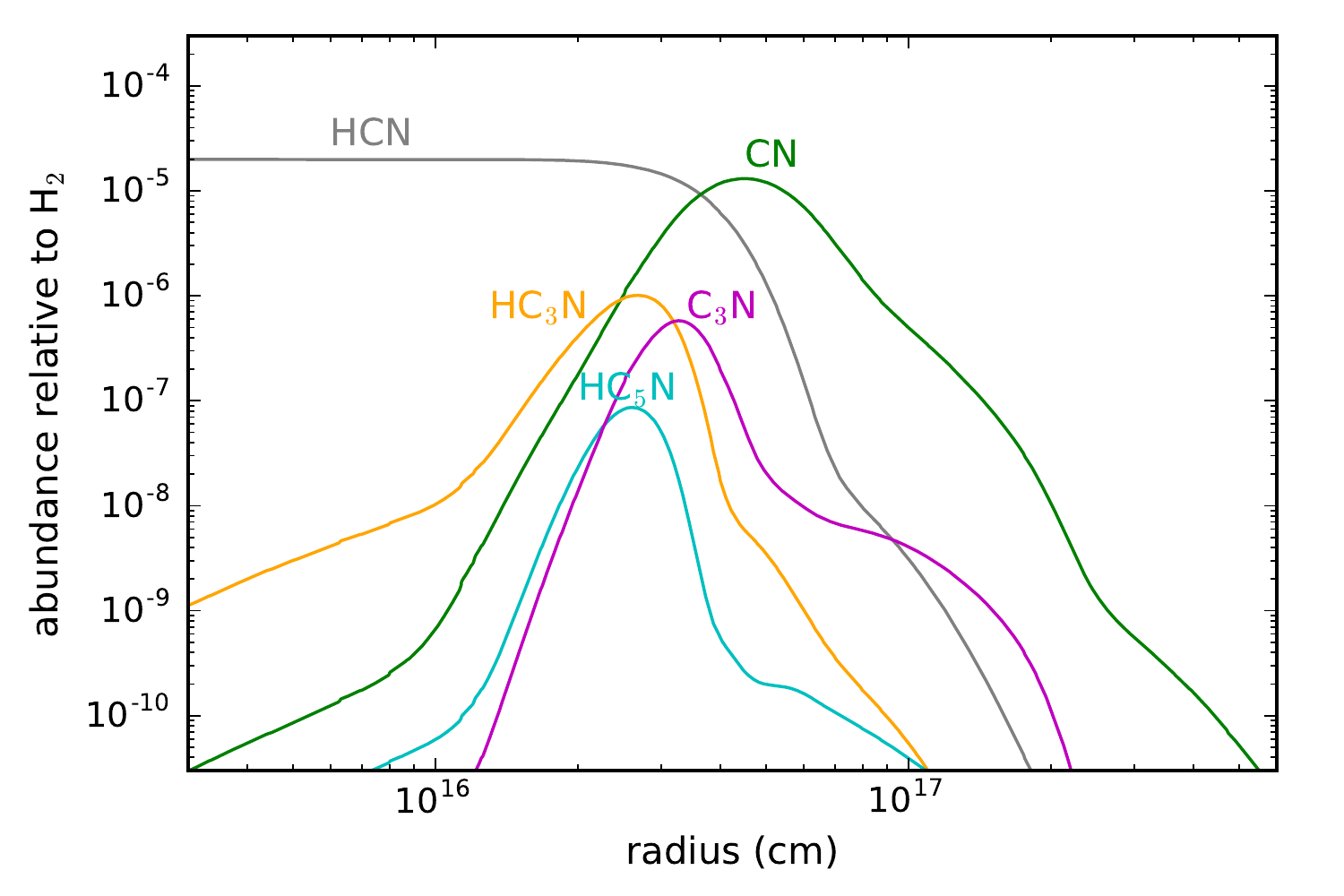}
\caption{Chemical schemes of growth of polyyne and cyanopolyyne carbon chains in IRC\,+10216 and calculated fractional abundances as a function of radius.} \label{fig:chemistry}
\end{figure*}

Fig.~\ref{fig:chemistry} summarizes some salient features of the chemical model in relation to the growth of carbon chains in IRC\,+10216. According to the model, the formation of carbon chains begins when the parent molecules C$_2$H$_2$ and HCN are photodissociated, in the region of the outer envelope where $A_V$$\sim$1 mag, producing the radicals C$_2$H and CN. A series of rapid neutral-neutral reactions involving these radicals then drives the growth of carbon chains. The schemes shown on top of Fig.~\ref{fig:chemistry}, inspired by \cite{mil1994} (1994), illustrate the main processes involved in the synthesis of polyynes and cyanopolyynes. The schemes include only those processes which, according to the chemical model, are the main contributors to the formation of each species. Nevertheless, a discussion of the reliability of these schemes and whether other types of reactions may play a role follows.

The formation of carbon chains begins with the photodissociation of C$_2$H$_2$ and HCN, whose cross sections are well known from laboratory experiments. We computed photodissociation rates for the local ultraviolet radiation field of \cite{dra1978} (1978) and dust shielding factors (see Table~\ref{table:rates}) using the {\footnotesize Meudon PDR} code (\cite{lep2006} 2006). For C$_2$H$_2$ we adopted the photoabsorption cross section measured by \cite{coo1995} (1995) and subtracted the photoionization cross section from the compilation by \cite{hud1971} (1971) to obtain the cross section of photodissociation, which is assumed to yield the C$_2$H radical with a 100~\% efficiency. For HCN, photodissociation is assumed to yield only CN radicals with the cross section measured by \cite{nut1982} (1982). Acetylene photodissociates faster than hydrogen cyanide, which translates into a shorter photodissociation radius for C$_2$H$_2$ (see bottom panels in Fig.~\ref{fig:chemistry}). The photodissociation rates computed here for C$_2$H$_2$ and HCN are similar to those given in the compilation of \cite{hea2017} (2017). These authors, however, compute a photodissociation rate of acetylene which is about twice smaller than our value, which would shift the photodissociation radius of C$_2$H$_2$ to somewhat outer radii with respect to our predictions.

\subsubsection{Polyynes}

The main pathway for forming polyynes of increasing chain length with an even number of carbon atoms involves successive reactions with the radical C$_2$H,
\begin{equation}
{\rm C_2H + C}_{2n}{\rm H_2} \rightarrow {\rm C}_{2n+2}{\rm H_2 + H}, \label{reac:c2h}
\end{equation}
where diacetylene (C$_4$H$_2$) is synthesized when $n=1$, triacetylene (C$_6$H$_2$) is formed when $n=2$, and so on. The reaction between C$_2$H and C$_2$H$_2$ is fast at low temperatures. The measured rate constant in the temperature range 15-295 K shows a slight negative temperature dependence, with values in the range $(1.1-2.3)\times10^{-10}$ cm$^3$ s$^{-1}$ (\cite{cha1998} 1998). There are no measurements of the rate constant at low temperatures for $n>1$. The reaction between C$_2$H and C$_4$H$_2$ has been studied theoretically and experimentally using the crossed beams technique (\cite{lan2008} 2008; \cite{gu2009} 2009). These studies indicate that the kinetics of the reaction at low temperatures should be similar to that of the reaction C$_2$H + C$_2$H$_2$, with a rate constant of the same order of magnitude and the H loss channel strongly favored. Reactions of C$_2$H with polyynes larger than C$_4$H$_2$ can be reasonably expected to proceed in a similar way, although there are no experimental or theoretical studies on the kinetics of reaction~(\ref{reac:c2h}) with $n>2$.

Triacetylene (C$_6$H$_2$) and larger polyynes are also formed by another route involving the C$_4$H radical,
\begin{equation}
{\rm C_4H + C}_{2n}{\rm H_2} \rightarrow {\rm C}_{2n+4}{\rm H_2 + H}, \label{reac:c4h}
\end{equation}
where the reaction of C$_4$H and C$_2$H$_2$ has been experimentally found to be rapid at low temperatures, with a rate constant of $(1.5-2.6)\times10^{-10}$ cm$^3$ s$^{-1}$ in the temperature range 39-300 K (\cite{ber2010} 2010). For polyynes larger than C$_6$H$_2$, routes similar to reactions~(\ref{reac:c2h}) and (\ref{reac:c4h}) but involving radicals larger than C$_4$H, such as C$_6$H and C$_8$H, are likely to show a similar kinetic behavior (see, e.g., \cite{sun2015} 2015) and thus to contribute significantly to the formation of large polyynes in environments such as IRC\,+10216.

Once the polyynes have been formed, the corresponding radicals, such as C$_4$H and C$_6$H, form directly by photodissociation, although reactions involving C$_2$ are also important but to a lesser extent. Diacetylene and triacetylene are easily photodissociated owing to their large absorption cross sections in the ultraviolet. The photodissociation rate of C$_4$H$_2$ has been computed from the photoabsorption cross section measured by \cite{fer2009} (2009), after subtraction of the photoionization cross section measured by \cite{sch2012} (2012). The main channel in the photodissociation of C$_4$H$_2$ is the production of C$_4$H radicals (\cite{sil2008} 2008). For C$_6$H$_2$, the photoionization cross section has not been measured and therefore the photodissociation rate has been directly computed from the photoabsorption cross section measured by \cite{klo1974} (1974) and \cite{shi2003} (2003), assuming that the main channel is C$_6$H + H.

The reactions involving C$_2$ of interest here are
\begin{equation}
{\rm C_2 + C}_{2n}{\rm H_2} \rightarrow {\rm C}_{2n+2}{\rm H + H}, \label{reac:c2}
\end{equation}
where the radicals C$_4$H and C$_6$H form by addition of C$_2$ subunits to C$_2$H$_2$ and C$_4$H$_2$, respectively. The reaction between C$_2$ and C$_2$H$_2$ is rapid at low temperatures, with a rate constant of $(2.4-4.8)\times10^{-10}$ cm$^3$ s$^{-1}$ in the temperature range 24-300 K (\cite{can2007} 2007; \cite{dau2008} 2008; \cite{par2008} 2008). The low-temperature kinetics of reactions of C$_2$ with C$_4$H$_2$ and larger polyynes has not been studied experimentally, and thus in the model we assume that they proceed with the same rate constant as the C$_2$ + C$_2$H$_2$ reaction.

According to the model, reactions~(\ref{reac:c2h}) and (\ref{reac:c4h}) account for the bulk of formation of polyynes in IRC\,+10216, while the radicals are mainly formed by photodissociation of the corresponding polyynes and by reaction~(\ref{reac:c2}). Other types of reactions, which are rapid and might be important synthetic routes in other environments, contribute little in IRC\,+10216 because they involve reactants which are not abundant enough in this source. For example, neutral carbon atoms and CH radicals have been experimentally found to react rapidly at low temperatures with C$_2$H$_2$ (\cite{cha2001} 2001; \cite{cla2002} 2002; \cite{can1997} 1997; \cite{loi2009} 2009) and they probably react rapidly with C$_4$H$_2$ and larger polyynes as well. These reactions tend to produce unsaturated carbon chains with an odd number of carbon atoms and in fact, in the chemical model, are major formation routes to species such as C$_3$H, C$_3$H$_2$, and C$_5$H, which are abundant in IRC\,+10216. However, here we focus on the chemistry of carbon chains C$_n$H and C$_n$H$_2$ with $n$ even, species for which reactions involving C atoms and CH radicals are not important. 

Molecular anions have been found to play a non-negligible role in the formation of neutral carbon chains in dark clouds (\cite{wal2009} 2009), and thus it is worth discussing whether they play a similar role in IRC\,+10216. According to the chemical model, reactions of the type
\begin{equation}
{\rm C}_{2n}{\rm H^- + H} \rightarrow {\rm C}_{2n}{\rm H_2 + e^-}, \label{reac:anions}
\end{equation}
which are rapid (\cite{eic2007} 2007), appear high in the list of reactions that contribute most to the formation of polyynes, especially at radii beyond $\sim$15$''$ from the star. However, since the precursor anion C$_{2n}$H$^-$ is mostly formed from its neutral counterpart C$_{2n}$H, which is in turn formed by photodissociation of the respective polyyne C$_{2n}$H$_2$, the net contribution of reaction~(\ref{reac:anions}) to the formation of polyynes and growth of carbon chains is essentially zero.

\subsubsection{Cyanopolyynes}

\begin{table*}
\caption{Important processes for the chemistry of carbon chains in IRC\,+10216} \label{table:rates} \centering
\begin{tabular}{l@{\hspace{0.1cm}}l@{\hspace{0.1cm}}lccccl}
\hline \hline
\multicolumn{3}{l}{Process} & \multicolumn{1}{c}{$\alpha$} &\multicolumn{1}{c}{$\beta$} & \multicolumn{1}{c}{$\gamma$} & \multicolumn{1}{c}{$\Delta T$ (K)} & \multicolumn{1}{l}{Notes and references} \\
\hline
\\
\multicolumn{6}{l}{Bimolecular gas-phase chemical reactions$^a$} \\
\hline
C$_2$H + C$_{2n}$H$_2$ & $\rightarrow$ & C$_{2n+2}$H$_2$ + H & $9.19\times10^{-11}$ & $-0.29$ & 0   & 15-295 & Based on C$_2$H + C$_2$H$_2$ (\cite{cha1998} 1998) \\
C$_4$H + C$_{2n}$H$_2$ & $\rightarrow$ & C$_{2n+4}$H$_2$ + H & $1.82\times10^{-10}$ & $-1.06$ & 66 & 39-300 & Based on C$_4$H + C$_2$H$_2$ (\cite{ber2010} 2010) \\
C$_2$ + C$_{2n}$H$_2$ & $\rightarrow$ & C$_{2n+2}$H + H           & $3.04\times10^{-10}$ & $-0.14$  & 0 & 24-300 & Based on C$_2$ + C$_2$H$_2$ (\cite{par2008} 2008) \\
CN + C$_{2n}$H$_2$ & $\rightarrow$ & HC$_{2n+1}$N + H             & $2.82\times10^{-10}$ & $-0.54$  & 22 & 25-295 & Based on CN + C$_2$H$_2$ (\cite{sim1993} 1993) \\
C$_3$N + C$_{2n}$H$_2$ & $\rightarrow$ & HC$_{2n+3}$N + H     & $3.21\times10^{-10}$ & $-0.12$  &  0 & 24-294 & Based on C$_3$N + C$_2$H$_2$ (\cite{fou2014} 2014) \\
\\
\multicolumn{6}{l}{Photodissociation processes$^b$} \\
\hline
C$_2$H$_2$ + $h\nu$ & $\rightarrow$ & C$_2$H + H                       & $4.40\times10^{-9}$ & & $2.46$  &  & \cite{coo1995} (1995) \\
C$_4$H$_2$ + $h\nu$ & $\rightarrow$ & C$_4$H + H                       & $1.09\times10^{-8}$ & & $2.19$  &  & \cite{fer2009} (2009); 100~\% C$_4$H + H (\cite{sil2008} 2008) \\
C$_6$H$_2$ + $h\nu$ & $\rightarrow$ & C$_6$H + H                       & $1.78\times10^{-8}$ & & $1.83$  &  & \cite{klo1974} (1974); \cite{shi2003} (2003) \\
HCN + $h\nu$ & $\rightarrow$ & CN + H                                            & $1.93\times10^{-9}$ & & $2.82$  &  & \cite{nut1982} (1982) \\
HC$_3$N + $h\nu$ & $\rightarrow$ & C$_3$N + H                            & $4.55\times10^{-9}$ & & $2.45$  &  & \cite{fer2009} (2009) \\
                                & $\rightarrow$ & C$_2$H + CN                          & $3.43\times10^{-9}$ & & $2.47$  &  & wavelength-dependent branching ratios (\cite{sil2009} 2009) \\
HC$_5$N + $h\nu$ & $\rightarrow$ & C$_5$N + H                            & $8.84\times10^{-9}$ & & $1.94$  &  & \cite{fra2010} (2010) \\
                                & $\rightarrow$ & C$_4$H + CN                          & $6.66\times10^{-9}$ & & $1.94$  &  & branching ratios of H and CN channels based on HC$_3$N \\
\hline
\end{tabular}
\tablenotea{$^a$ Rate constants have units of cm$^3$ s$^{-1}$ and are given by the expression $k(T) = \alpha (T/300)^\beta \exp (-\gamma/T)$. $\Delta T$ is the valid temperature range of the rate constant expression. The subindex $n$ has values 1, 2, 3, 4, ....\\
$^b$ Photodissociation rates have units of s$^{-1}$ and are given by the expression $K = \alpha \exp (-\gamma A_V)$.}
\end{table*}

Cyanopolyynes in IRC\,+10216 are formed in a bottom-up scenario analogous with polyynes (see scheme on the top-right of Fig.\ref{fig:chemistry}). We first concentrate on the potentially important synthetic route given by the reaction
\begin{equation}
{\rm C}_{2n}{\rm H + HC}_{2m-1}{\rm N} \rightarrow {\rm HC}_{2n+2m-1}{\rm N + H}, \label{reac:c2h_hcn}
\end{equation}
where the simplest case is with $n=1$ and $m=1$, which corresponds to the reaction between C$_2$H and HCN to form HC$_3$N. The rate constant of this reaction has not been measured at low temperatures but measurements in the temperature range 297-360 K (\cite{hoo1997} 1997) and theoretical studies (\cite{fuk1997} 1997) point to the presence of an activation barrier of $\sim$800 K that would make the reaction too slow at low temperatures. For this reason, we assume that reaction~(\ref{reac:c2h_hcn}) shows the same kinetic behavior for values of $n$ and/or $m$ higher than one, i.e., for reactions of C$_2$H, C$_4$H, or larger radicals with HCN, HC$_3$N, or larger cyanopolyynes. As a consequence, reaction~(\ref{reac:c2h_hcn}) is not efficient in IRC\,+10216, and is the reason it does not appear in the scheme on the top-right of Fig.~\ref{fig:chemistry}. The UMIST and KIDA databases make different assumptions on the kinetic behavior of reaction~(\ref{reac:c2h_hcn}) for $n$ and/or $m$ $>$1. For example, the UMIST database does not consider these reactions while in the KIDA database some of these, e.g., the reaction C$_2$H + HC$_3$N, are assumed to be rapid at low temperatures. It would be of great interest to study the low-temperature chemical kinetics of reactions such as C$_2$H + HC$_3$N and C$_4$H + HCN to clarify whether there are some values of $n$ and $m$ for which reaction~(\ref{reac:c2h_hcn}) is rapid at low temperatures, and thus contributes to the growth of cyanopolyynes in IRC\,+10216 and in cold interstellar clouds.

If reaction~(\ref{reac:c2h_hcn}) is indeed completely closed at low temperatures, then the growth of cyanopolyynes does not occur at the expense of cyanopolyynes of lower size but at the expense of polyynes in reactions involving the CN radical
\begin{equation}
{\rm CN + C}_{2n}{\rm H_2} \rightarrow {\rm HC}_{2n+1}{\rm N + H}, \label{reac:cn}
\end{equation}
where C$_2$H$_2$ serves as the precursor of HC$_3$N, the polyyne C$_4$H$_2$ for HC$_5$N, and so on. Reaction~(\ref{reac:cn}) with $n = 1$, i.e., CN + C$_2$H$_2$, has been experimentally found to be fast at low temperatures, with a rate constant of $(2.6-4.7)\times10^{-10}$ cm$^3$ s$^{-1}$ in the temperature range 25-298 K (\cite{sim1993} 1993). Other studies confirm that the reaction proceeds without an activation barrier yielding HC$_3$N as the main product (\cite{fuk1997} 1997; \cite{hua2000} 2000; \cite{cho2004} 2004). Reaction~(\ref{reac:cn}) with $n = 2$, i.e., CN + C$_4$H$_2$, is also rapid at room temperature, with a measured rate constant of $4.2\times10^{-10}$ cm$^3$ s$^{-1}$ (\cite{sek1996a} 1996a), although it has not been experimentally studied at low temperatures. Theoretical calculations and crossed beams experiments, however, indicate that the reaction of CN with C$_4$H$_2$ as well as those of CN with larger polyynes are rapid at low temperatures and provide an efficient route to large cyanopolyynes (\cite{fuk1998} 1998; \cite{zha2009} 2009). According to our chemical model, reaction~(\ref{reac:cn}) serves as the main formation route to cyanopolyynes in IRC\,+10216.

For cyanopolyynes larger than HC$_3$N, there is another important formation route involving the radical C$_3$N
\begin{equation}
{\rm C_3N + C}_{2n}{\rm H_2} \rightarrow {\rm HC}_{2n+3}{\rm N + H}. \label{reac:c3n}
\end{equation}
Indeed it has been recently found that the radical C$_3$N reacts as fast as the radical CN with C$_2$H$_2$, with a rate constant of $(2.5-4.7)\times10^{-10}$ cm$^3$ s$^{-1}$ in the temperature range 24-294 K (\cite{fou2014} 2014). Assuming that the same kinetic behavior holds for C$_3$N reacting with C$_4$H$_2$ and longer polyynes, reaction~(\ref{reac:c3n}) becomes an important formation route of cyanopolyynes. In fact, in our chemical model reactions~(\ref{reac:cn}) and (\ref{reac:c3n}) contribute similarly to the formation of HC$_5$N. Moreover, routes similar to reactions~(\ref{reac:cn}) and (\ref{reac:c3n}) but involving radicals larger than C$_3$N, such as C$_5$N, are likely significant contributors to the formation of cyanopolyynes larger than HC$_5$N in IRC\,+10216.

The radicals C$_3$N and C$_5$N are mainly formed by direct photodissociation of the corresponding cyanopolyynes, HC$_3$N and HC$_5$N, respectively. As with polyynes, cyanopolyynes have large absorption cross sections at ultraviolet wavelengths and are thus easily photodissociated. The photodissociation rate of HC$_3$N has been computed from the photoabsorption cross section measured by \cite{fer2009} (2009), after subtraction of the photoionization cross section measured by \cite{lea2014} (2014). We have adopted the wavelength-dependent branching ratios computed by \cite{sil2009} (2009) for the two open channels, C$_3$N + H and C$_2$H + CN, which are in agreement with experiments (\cite{hal1988} 1988; \cite{cla1995} 1995; \cite{sek1996b} 1996b). In the case of HC$_5$N, the ionization potential and photoionization cross section are not known, and therefore the photodissociation rate has been computed from the photoabsorption cross section measured by \cite{fra2010} (2010), where we have assumed that the H and CN channels occur with similar branching ratios as for HC$_3$N.

\subsection{Radiative transfer models: emission of carbon chains} \label{sec:radtra_model}

\begin{figure*}
\centering
\includegraphics[angle=0,width=\columnwidth]{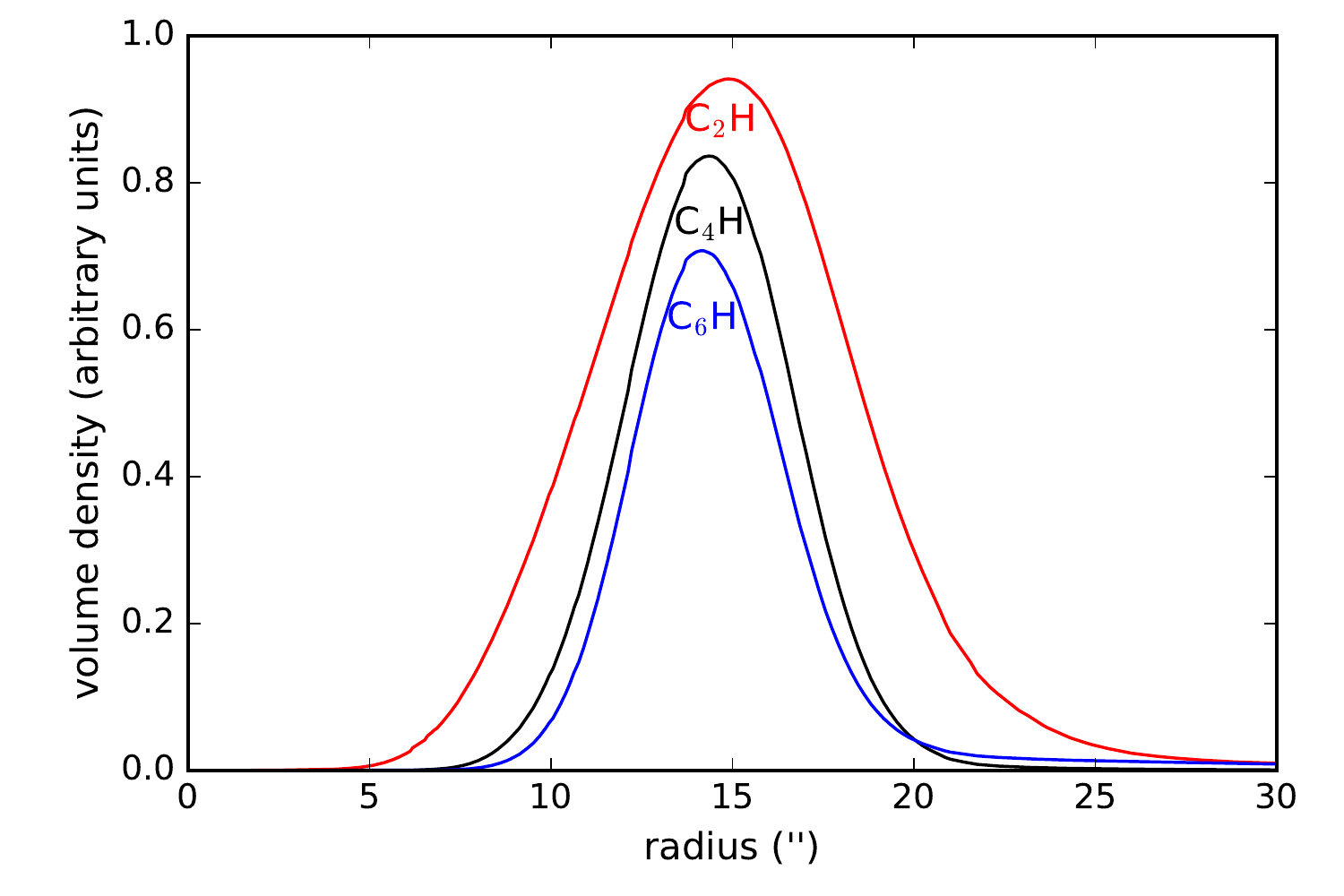}
\includegraphics[angle=0,width=\columnwidth]{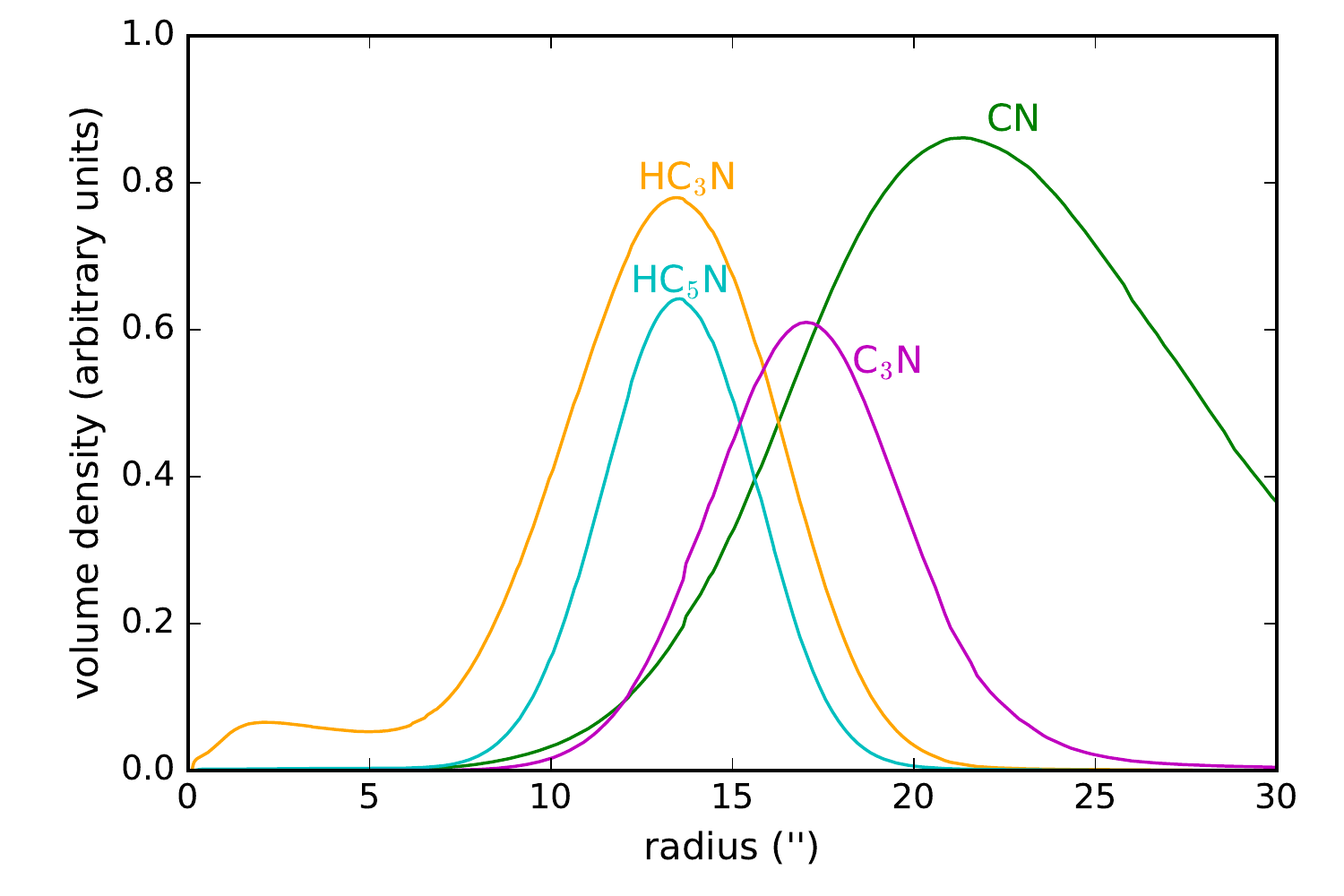}
\includegraphics[angle=0,width=\columnwidth]{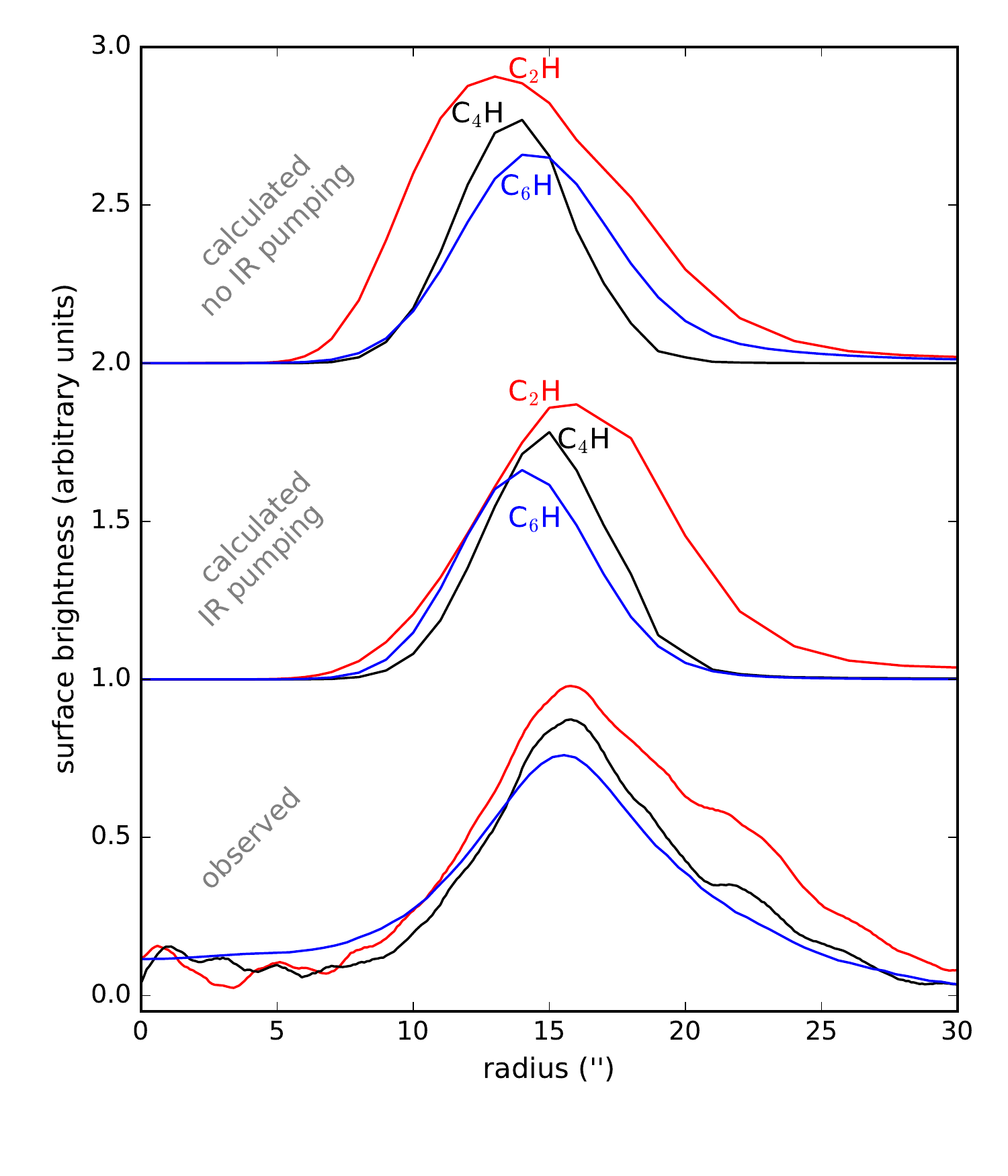}
\includegraphics[angle=0,width=\columnwidth]{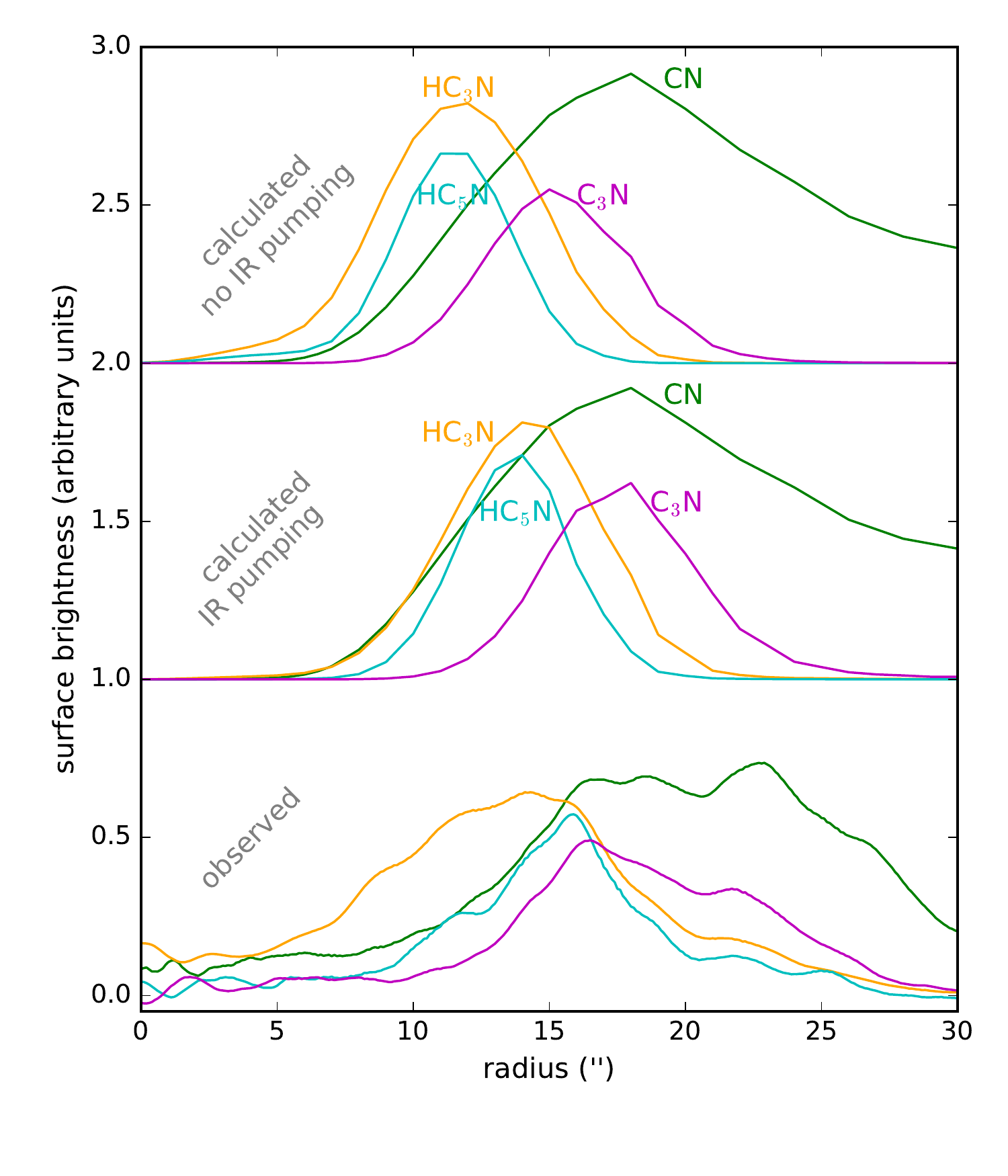}
\caption{\emph{Top panels.--} Calculated radial distributions of abundances expressed as number of particles per unit volume. Curves have been arbitrarily scaled in the $y$-axis for display. \emph{Bottom panels.--} Comparison between calculated (neglecting and including infrared pumping) and observed radial distributions of the emission at velocities around $V_{\rm LSR}=V_{\rm sys}$ for the lines of C$_2$H, C$_4$H, and C$_6$H (\emph{left}), and CN, C$_3$N, HC$_3$N, and HC$_5$N (\emph{right}) plotted in Fig.~\ref{fig:map_all}. The observed brightness distributions are those previously shown in Fig.~\ref{fig:azimuth_average}. Curves have been arbitrarily scaled and shifted along the $y$-axis for display.} \label{fig:distribution}
\end{figure*}

In the top panels of Fig.~\ref{fig:distribution}, the abundance radial profiles calculated for C$_2$H, C$_4$H, and C$_6$H (left), and for CN, HC$_3$N, HC$_5$N, and C$_3$N (right) are shown, in which abundances are expressed in absolute rather than relative terms, i.e., as number of particles per unit volume. These abundance distributions provide a good indication of how the rotational emission of each species is distributed as a function of radius in the plane of the sky. Nevertheless, significant differences can arise depending on the transition owing to excitation effects, because of the way the rotational levels are populated through collisions and radiative processes such as infrared pumping at different radii. To take into account these effects we have carried out excitation and radiative transfer calculations to convert the abundance radial distributions into radial brightness distributions of $\lambda$~3~mm transitions, which can in turn be compared with the ALMA observations.

The radiative transfer calculations have been performed with the multi-shell LVG code and physical model of IRC\,+10216 used by \cite{agu2012} (2012), with the radial abundance distributions calculated with the chemical model. The calculations require spectroscopic and collision excitation data, which have different degrees of accuracy depending on the species and type of data. Rotational constants of the ground vibrational state, and for some species, vibrationally excited states as well, are well known from microwave laboratory experiments: C$_2$H (\cite{mul2000} 2000; \cite{kil2007} 2007), CN (\cite{kli1995} 1995), C$_4$H and C$_3$N (\cite{got1983} 1983), C$_6$H (\cite{lin1999} 1999), HC$_3$N (\cite{tho2000} 2000), and HC$_5$N (\cite{biz2004} 2004). Dipole moments have been measured for the radical CN (\cite{tho1968} 1968) and the closed electronic-shell species HC$_3$N (\cite{del1985} 1985) and HC$_5$N (\cite{ale1976} 1976), and ab initio theoretical values, with significantly higher uncertainties, have been calculated for the radicals C$_2$H, C$_4$H, C$_6$H (\cite{woo1995} 1995), and C$_3$N (\cite{mcc1995} 1995).

In addition to thermal excitation via collisions with H$_2$ and He, absorption of infrared photons and pumping to excited vibrational states, followed by radiative decay to rotational levels in the ground vibrational state, is an important excitation mechanism of molecules in IRC\,+10216 (\cite{deg1984} 1984; \cite{agu2006} 2006; \cite{gon2007} 2007; \cite{agu2008} 2008, 2015; \cite{cor2009} 2009; \cite{dan2012} 2012; \cite{deb2012} 2012). Here, we have included excitation through infrared pumping for all studied species, mostly through bands lying in the mid infrared, where the flux in IRC\,+10216 is large (\cite{cer1999} 1999). In the case of the radicals, to facilitate the excitation and radiative transfer calculations, we have collapsed the fine rotational structure and simply treated these species as linear molecules with a $^1\Sigma$ electronic state. For C$_2$H, we have included the first four vibrationally excited states of the bending mode ($\nu_2$ = 1, 2, 3, 4), and the first vibrationally excited states of the stretching modes ($\nu_1$ = 1 and $\nu_3$ = 1). The vibrationally excited state that plays the most important role, via infrared pumping, in the excitation of C$_2$H in IRC\,+10216 is $\nu_2$ = 1, which lies 371 cm$^{-1}$ above the ground vibrational state. The wavelengths and strengths of the vibrational bands have been taken from \cite{tar2004} (2004). For the radical CN, we have included the $v$ = 0 $\rightarrow$ 1 band, lying at 2042 cm$^{-1}$ (\cite{hub2005} 2005; \cite{bro2014} 2014), which however plays a minor role on the excitation of the $\lambda$~3~mm lines in IRC\,+10216. For HC$_3$N, we have included the first excited states of the vibrational bending modes $\nu_5$ and $\nu_6$, which have strong fundamental bands at 663 and 498 cm$^{-1}$. The wavelengths and strengths of the vibrational bands are from the compilation by J. Crovisier\footnote{\texttt{www.lesia.obspm.fr/perso/jacques-crovisier/basemole}}, which are based on extensive laboratory work (e.g., \cite{uye1982} 1982; \cite{jol2007} 2007). For cyanodiacetylene, we have included the first excited states of the vibrational bending modes $\nu_7$ and $\nu_8$, whose calculated fundamental bands, lying at 566 and 685 cm$^{-1}$, have been found to be important for the rotational excitation of HC$_5$N in IRC\,+10216 (\cite{deg1984} 1984). For the radicals C$_4$H, C$_6$H, and C$_3$N there is little information on the wavelengths and strengths of vibrational bands. For these species we have instead included a generic vibrationally excited state lying at 15 $\mu$m above the ground vibrational state, with an Einstein coefficient of spontaneous emission of 5 s$^{-1}$ for the P(1) transition of the vibrational band. A similar treatment, with slightly different parameters, was adopted for C$_4$H and C$_6$H by \cite{cor2009} (2009).

Rate coefficients for rotational excitation induced by collisions with H$_2$ and He are scarce in the literature for the species studied here. We have adopted the rate coefficients calculated by \cite{dum2010} (2010) for HCN in collisions with He, properly corrected by the square root of the ratio of reduced masses with He and H$_2$, for the excitation of the radicals C$_2$H and CN. \cite{gre1978} (1978) calculated rate coefficients for the excitation of HC$_3$N in collisions with He, and those values, corrected for H$_2$ and extrapolated to higher $J$ levels through the IOS approximation when needed, have been used for HC$_3$N as well as for the radicals C$_3$N and C$_4$H. For the longer carbon chains HC$_5$N and C$_6$H, we have adopted collision rate coefficients from the approximate generic expression given by \cite{deg1984} (1984).

The calculated radial distributions of brightness in the plane of the sky are shown in the bottom panels of Fig.~\ref{fig:distribution}, in comparison with the observed ones. We leave the discussion of the comparison between calculated and observed emission distributions for Section~\ref{sec:discussion} and focus here on the comparison between radial distributions of abundance and emission (top panels and upper and middle curves in bottom panels of Fig.~\ref{fig:distribution}). When infrared pumping is neglected and rotational excitation occurs exclusively through collisions with H$_2$ and He (upper curves in bottom panels of Fig.~\ref{fig:distribution}), calculations show that the radius where the $\lambda$~3~mm emission is maximum does not exactly coincide with where the number of particles per unit volume is maximum. In general, the emission tends to peak a few arcseconds inward from where the abundance is maximum. This is evident in C$_2$H, CN, C$_3$N, and the cyanopolyynes HC$_3$N and HC$_5$N. The origin of this effect is that the circumstellar envelope has strong radial gradients of density and temperature. More internal regions are characterized by denser and warmer gas, and the excitation of rotational levels by collisions is more favorable. Thus, the radius of maximum brightness represents a compromise between a high absolute abundance and a high excitation rate through collisions. The shift to shorter radii of the maximum of brightness becomes more pronounced for high-$J$ transitions, whose upper levels are high in energy and are more difficult to excite through collisions in the lower density environments.

The situation may change significantly when infrared pumping to excited vibrational states is included in the calculations (middle curves in bottom panels of Fig.~\ref{fig:distribution}). If a species has strong vibrational bands at wavelengths at which the flux of infrared photons in the envelope is high, then infrared pumping may dominate over collisions in exciting the rotational levels of the ground vibrational state. Our calculations indicate that for most of the species studied here, infrared pumping plays an important role in the excitation of the $\lambda$~3~mm lines, and in some cases has important consequences for the emerging line intensities and radial distributions of the brightness. Infrared pumping tends to increase the excitation temperatures of the rotational transitions of the ground vibrational state and the intensities of the rotational lines, and extends the emission region to outer locations in the envelope. The importance of this effect, however, varies depending on the species and transition. Among the species studied here, infrared pumping has a marked effect on the brightness radial distribution of the $\lambda$~3~mm lines of C$_2$H (through absorption of 27~$\mu$m photons and pumping to the $\nu_2$ = 1 state), HC$_3$N (by excitation of the $\nu_5$ = 1 and $\nu_6$ = 1 states with 15~$\mu$m and 20~$\mu$m photons), and HC$_5$N (exciting the $\nu_7$ = 1 and $\nu_8$ = 1 states with 14.6 and 17.7~$\mu$m photons). In contrast, there is little effect of infrared pumping on the emission distribution of the $N=1-0$ line of CN, which can only be pumped by 4.9 $\mu$m photons through the fundamental band of the stretching mode of the C--N bond. In the case of the radicals C$_4$H, C$_6$H, and C$_3$N, the hypothesized vibrational band at 15~$\mu$m has different effects on the $\lambda$~3~mm emission distribution of each species, the effects being more marked for C$_3$N, somewhat less important for C$_4$H, and of little importance for C$_6$H. We however would like to stress that the lack of knowledge of the wavelengths and strengths of vibrational bands for these three radicals create significant uncertainties in the predicted brightness radial distributions. Moreover, the presence of density enhancements in the form of arcs, which are not included in the current model, will increase the importance of collisions with respect to infrared pumping, modifying the relative roles of both processes in the excitation of the $\lambda$~3~mm lines.

In summary, although the abundance radial distribution provides a good, first-order estimate of how the $\lambda$~3~mm emission is distributed radially, there can be shifts of up to several arcseconds depending on the peculiarities of the excitation of each species.

\section{Discussion} \label{sec:discussion}

A comparison between the observed and calculated $\lambda$~3~mm emission distributions of carbon chains allows us to draw conclusions about the formation mechanism of carbon chains in IRC\,+10216. For this discussion we focus on Fig.~\ref{fig:distribution}, specifically on the bottom panels, where calculated and observed brightness radial distributions are compared. We concentrate on the emission distributions calculated when infrared pumping is taken into account (middle curves in bottom panels of fig.~\ref{fig:distribution}), which are more realistic than those obtained when excitation by collisions is only considered (upper curves in bottom panels of fig.~\ref{fig:distribution}), even if just an approximate treatment of the pumping scheme is adopted, as occurs for the radicals C$_4$H, C$_6$H, and C$_3$N. Before drawing firm conclusions, we must keep in mind that the observed brightness radial distributions are the result of an azimuthal average of the brightness distributions shown in Fig.~\ref{fig:map_all}. These maps show the existence of an azimuthal structure, with shells that are not strictly concentric, as indicated by the presence of inter-crossing arcs, probably the consequence of a mass loss process that is not strictly continuous and isotropic. Because the envelope in our model is the result of an isotropic uniform mass loss process, and the angular resolution has been degraded in the observational data owing to the azimuthal average, the comparison between the calculated and observed emission distributions can only be done to first order, and not at the sub-arcsecond level.

Observations locate the emission of the three hydrocarbon radicals C$_2$H, C$_4$H, and C$_6$H at the same radial position in the envelope, with maxima between $15\secp5$ and 16$''$ from the star. It is remarkable that the chemical model also predicts a similar behavior of the abundance radial profiles, with a slight radial stratification in which the radial position decreases as the size of the radical increases (see top-left panel in Fig.~\ref{fig:distribution}). This radial stratification is slightly more marked when abundances are converted into brightness distributions (middle curves in bottom-right panel of Fig.~\ref{fig:distribution}), although the predictions for C$_4$H and C$_6$H have to be viewed with caution due to the approximate treatment of infrared pumping. Taking into account that the formation of these radicals in the chemical model occurs sequentially with respect to increasing size (see scheme on the top-right of Fig.~\ref{fig:chemistry}), one might intuitively expect that the time delay needed to produce an increase in the size of the carbon chain (e.g., to go from C$_2$H to C$_4$H via the reaction C$_2$H + C$_2$H$_2$ followed by the photodissociation of C$_4$H$_2$) should result in a radial shift, in which the larger carbon chains should appear later in the expansion, and thus farther from the star. Such behaviour is neither predicted by the chemical model nor observed in the maps, which is a consequence of the rapidity at which photochemistry takes place compared with the expansion of the envelope in IRC\,+10216. This finding is important, since it limits the use of the brightness radial distributions of C$_2$H, C$_4$H, and C$_6$H as a chemical clock. The chemical model predicts that a significant radial stratification with hydrocarbon radicals C$_{2n}$H of increasing size would take place only if the chemical reactions were significantly slower or the expanding wind were significantly faster. We stress that this conclusion is reached using numerical simulations to model the circumstellar chemistry, and that simple timescale arguments may lead to erroneous conclusions.

The CN-bearing species show more diverse brightness radial distributions than the hydrocarbon radicals as a result of a slightly different chemistry. We first note that the emission of the radical CN shows a rather broad radial distribution, extending to radii greater than 30$''$. This observational fact is correctly predicted by the chemical and excitation model. The slower photodissociation rate of HCN, compared to that of C$_2$H$_2$, is at the origin of the much more extended radial distribution of CN emission with respect to that of C$_2$H. A second interesting aspect is that the observed emission of HC$_3$N appears at rather short radial distances (i.e., inner to 10$''$), a fact that is fairly well reproduced by the model. Such an early appearance is related to the fact that HC$_3$N is mainly formed in the reaction between C$_2$H$_2$, a parent molecule which is injected from the innermost circumstellar regions and is destroyed by photodissociation in the outer envelope, and CN, a daughter species that is formed by photodissociation of HCN. Therefore, HC$_3$N must be formed when the abundances of both reactants, having nearly complementary distributions, are high enough. This situation occurs in a transition region of the envelope where the abundance of acetylene decreases as the gas expands while that of CN increases. Thus, species whose direct precursor is a parent species may be considered as first-generation daughter species as opposed to second-generation daughter species, which are instead formed from precursors that are themselves daughter species. First-generation daughter species, such as HC$_3$N, thus tend to appear earlier in the expansion than second-generation daughter species, such as the radicals C$_4$H, C$_6$H, and C$_3$N. The observed emission of the longer cyanopolyyne HC$_5$N rises at larger radii than that of HC$_3$N, an observational fact that is not reproduced by the model, which instead predicts similar radial distributions for both cyanopolyynes in terms of both the abundance and the $\lambda$~3~mm emission. The early appearance of HC$_5$N in the model can be traced to the important contribution of the formation route involving the C$_3$N + C$_2$H$_2$ reaction, which involves as reactant the parent species C$_2$H$_2$ and thus occurs in inner regions than the other important HC$_5$N-forming reaction CN + C$_4$H$_2$. Stated differently, the model predicts for HC$_5$N a more marked character of first-generation daughter species than suggested by observations. The discrepancy for HC$_5$N is more likely due to the chemical model rather than to the excitation and radiative transfer calculations.

An interesting piece of information is provided by the availability of observational data for the electronic closed-shell molecule HC$_3$N and its photodissociation descendant C$_3$N. The observations show that the brightness distribution of C$_3$N is shifted to outer radii with respect to that of HC$_3$N (see bottom-right panel in Fig.~\ref{fig:distribution}), a result that is well reproduced by the current model. This result suggests that analogous pairs of species, whose spatial distribution may be more difficult to map, may behave similarly as, for example, occurs for the pair HC$_5$N/C$_5$N, in which the rotational lines of the radical are too weak to map, or the pairs C$_4$H$_2$/C$_4$H and C$_6$H$_2$/C$_6$H, where the polyynes lack a permanent dipole moment.

A last aspect that is worth discussing concerns the fact that the model assumes a density radial distribution that decreases smoothly with increasing radius, while observations show that a good fraction of the circumstellar matter is locked into clumpy arcs. The non-uniform density structure can affect the abundance and emission radial distributions of the molecules in different ways. First, external ultraviolet photons can penetrate more easily through the regions between clumps where the density of gas and dust is depressed, allowing photochemical processes to take place inside the envelope (\cite{dec2010} 2010; \cite{agu2010} 2010). This effect can explain the formation in inner circumstellar regions of some key molecules such as the hydrides H$_2$O and NH$_3$, for which conventional chemical models are not able to account for their formation. In contrast, our ALMA observations show that carbon chains, which are efficiently produced by photochemistry, are confined to the outer envelope, with no or little presence in inner regions. The action of photochemistry in the inner envelope must therefore affect only a few selected regions or be selective with respect to the formation of different types of molecules owing to, for example, temperature effects (warm photochemistry in inner regions vs. cold photochemistry in the outer envelope). Second, the existence of shells and arcs where the density is enhanced over the surrounding medium may provide shelters to carbon chains, because the enhanced densities boost the rates of carbon chain-forming reactions and provide a more efficient shielding against photodestruction. Higher densities also increase the rates of rotational excitation by collisions with H$_2$ and He. Density-enhanced shells may therefore provide environments where both the fractional abundance and the emission in rotational lines of carbon chains are enhanced over the surrounding medium. These effects have been studied by \cite{bro2003} (2003) and more recently by \cite{cor2009} (2009) in a spherical model of the envelope with shells located at specific radii where densities are enhanced over the surrounding circumstellar medium. The model predicts that the radial distribution of carbon chains narrows and the rotational emission is mainly confined to the shells of enhanced density. Some features observed with the IRAM and VLA interferometers (\cite{gue1997} 1997, 1999; \cite{din2008} 2008) are reproduced by this model, but there are important discrepancies for species such as C$_4$H. In view of the rich azimuthal structure revealed by the ALMA maps, next generation chemical models of IRC\,+10216 will probably need to relax the spherical assumption and to either study the chemistry radially at different PA in the plane of the sky or move to a three-dimensional model of the envelope that takes into account the entire structure of shells and arcs which make up the circumstellar envelope.

\section{Concluding remarks}

We have carried out a combined observational and modeling study to understand the growth of carbon chains in the C-star envelope IRC\,+10216. We have used ALMA to map at high sensitivity and high angular resolution the emission of $\lambda$~3~mm rotational lines of the hydrocarbon radicals C$_2$H, C$_4$H, and C$_6$H, and the CN-containing species CN, C$_3$N, HC$_3$N, and HC$_5$N. All of these species are distributed in a hollow spherical shell of width 5-10$''$ located at a distance of 10-20$''$ from the star, and none show appreciable compact emission around the star. The broad 5-10$''$ wide hollow shell seen at an angular resolution of $\sim$1$''$ shows various thin shells of width 1-2$''$, which are not strictly concentric, pointing to a discontinuous and non-isotropic mass loss process that could be driven by the presence of a companion star. The hydrocarbon radicals C$_2$H, C$_4$H, and C$_6$H show very similar radial distributions, while the CN-containing species show a radial stratification, with HC$_3$N appearing at shorter radii and the CN radical extending out to larger radii. This behavior is reasonably well reproduced by our chemical model, which validates a scenario of carbon chain growth initiated by photodissociation of acetylene and hydrogen cyanide, rapid gas-phase chemical reactions of C$_2$H and C$_4$H radicals with unsaturated hydrocarbons as the main mechanism responsible for the growth of polyynes, and reactions of CN and C$_3$N radicals with unsaturated hydrocarbons as the dominant formation routes to cyanopolyynes.

\acknowledgements

We acknowledge funding support from the European Research Council (ERC Grant 610256: NANOCOSMOS) and from Spanish MINECO through grants AYA2012-32032 and AYA2016-75066-C2-1-P. M.A. also thanks funding support from the Ram\'on y Cajal programme of Spanish MINECO (RyC-2014-16277).

\clearpage

\begin{appendix}

\section{Emission maps in different rotational transitions of C$_4$H, C$_3$N, HC$_3$N, and HC$_5$N} \label{sec:map_trans}

\begin{figure}[b!]
\centering
\includegraphics[angle=0,width=\columnwidth]{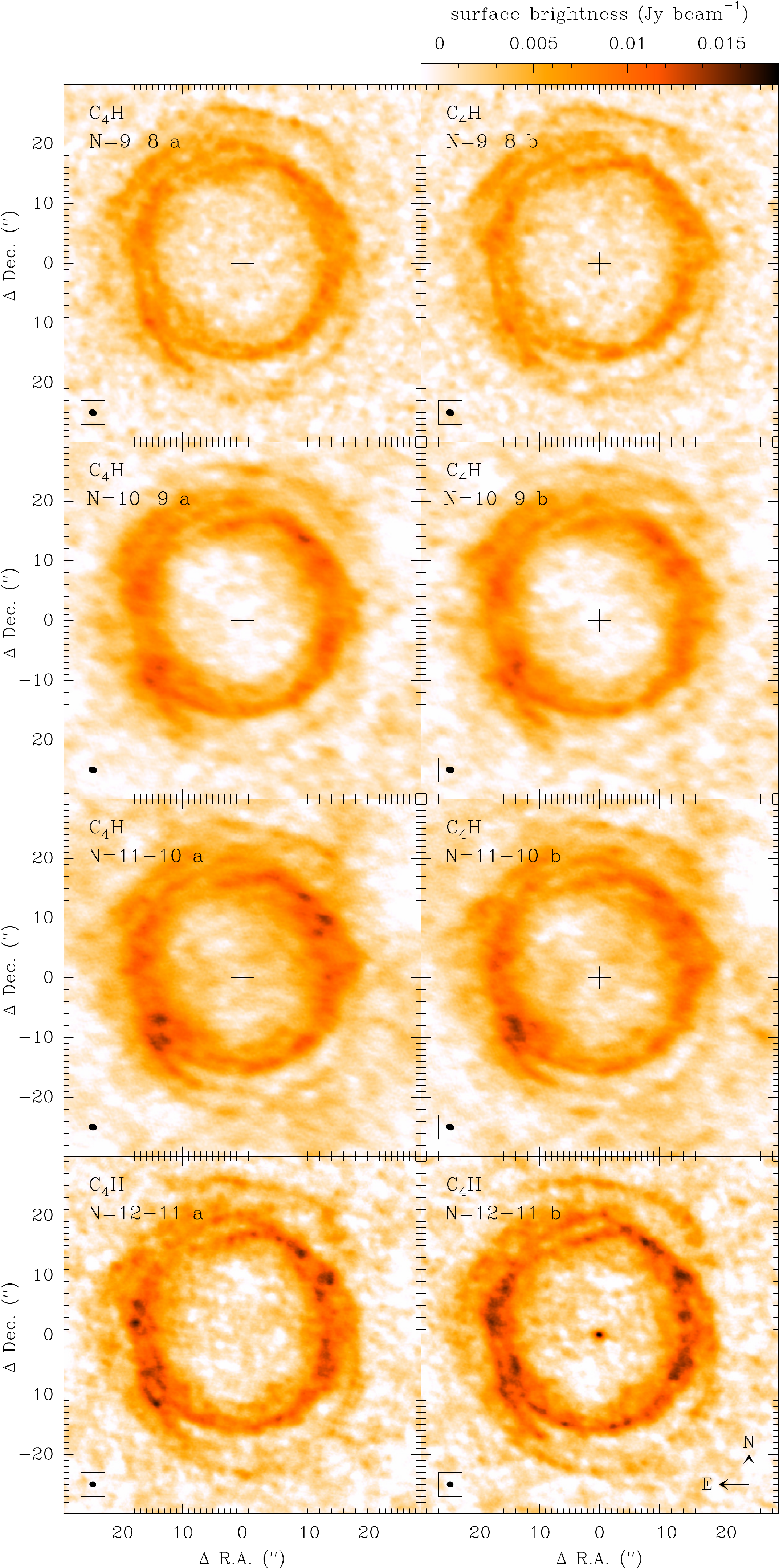}
\caption{Brightness distributions of the $N=9-8$ through $N=12-11$ spin-rotation doublets of C$_4$H, averaged over the central 3 km s$^{-1}$ of each line. The size ($\sim1''$ for all lines) and shape of the synthesized beam is shown in the bottom left corner of each panel. Maps are centered on the position of the star, indicated by a cross. The compact unresolved emission centered on the star in the panel corresponding to the C$_4$H $N=12-11$ b line arises from the $v=1$ $J=1-0$ transition of CO.} \label{fig:map_c4h}
\end{figure}

\begin{figure}[b!]
\centering
\includegraphics[angle=0,width=\columnwidth]{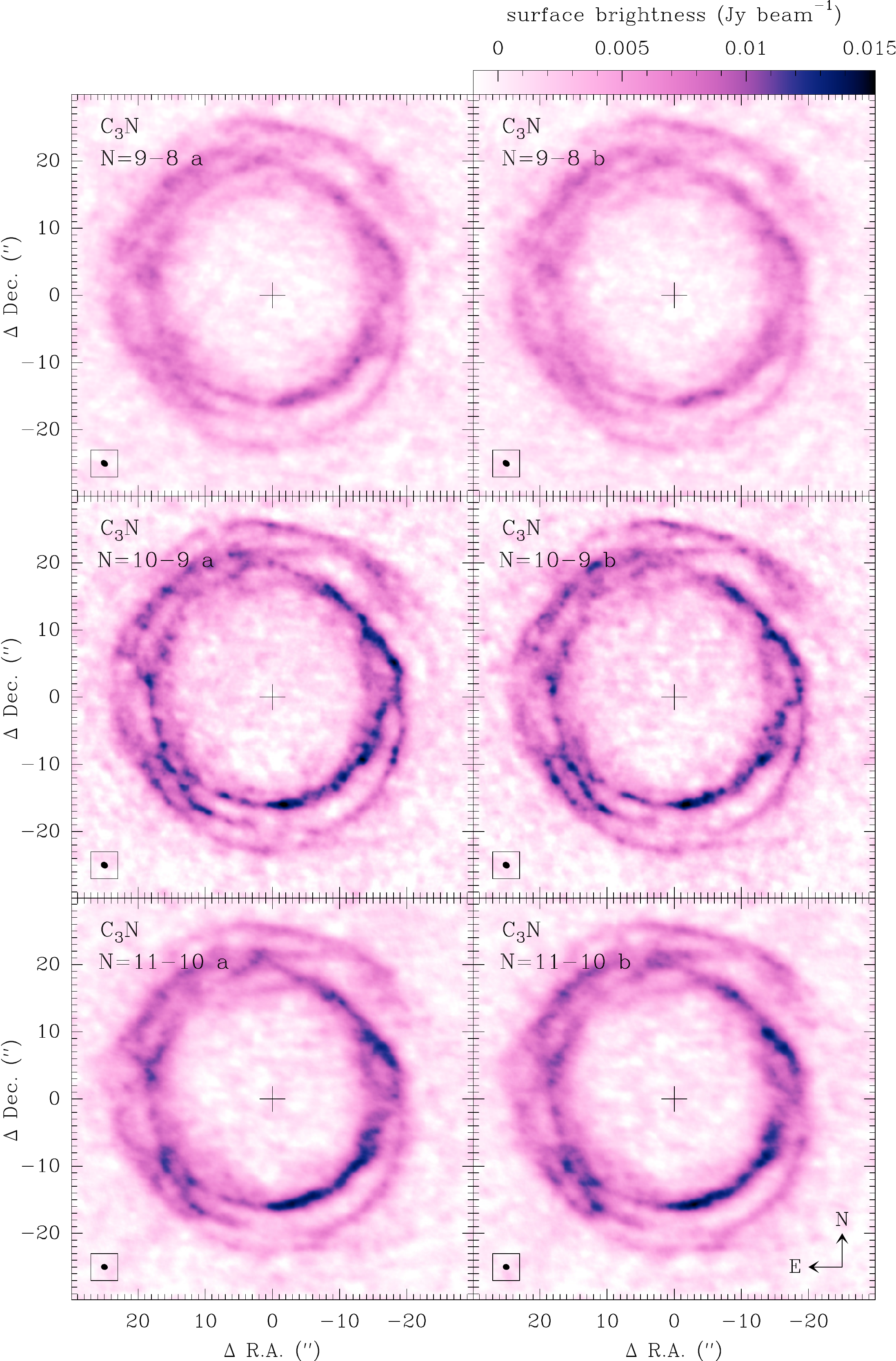}
\caption{Brightness distributions of the $N=9-8$ through $N=11-10$ spin-rotation doublets of C$_3$N, averaged over the central 3 km s$^{-1}$ of each line. The size ($\sim1''$ for all lines) and shape of the synthesized beam is shown in the bottom left corner of each panel. Maps are centered on the position of the star, indicated by a cross.} \label{fig:map_c3n}
\end{figure}

\begin{figure}
\centering
\includegraphics[angle=0,width=\columnwidth]{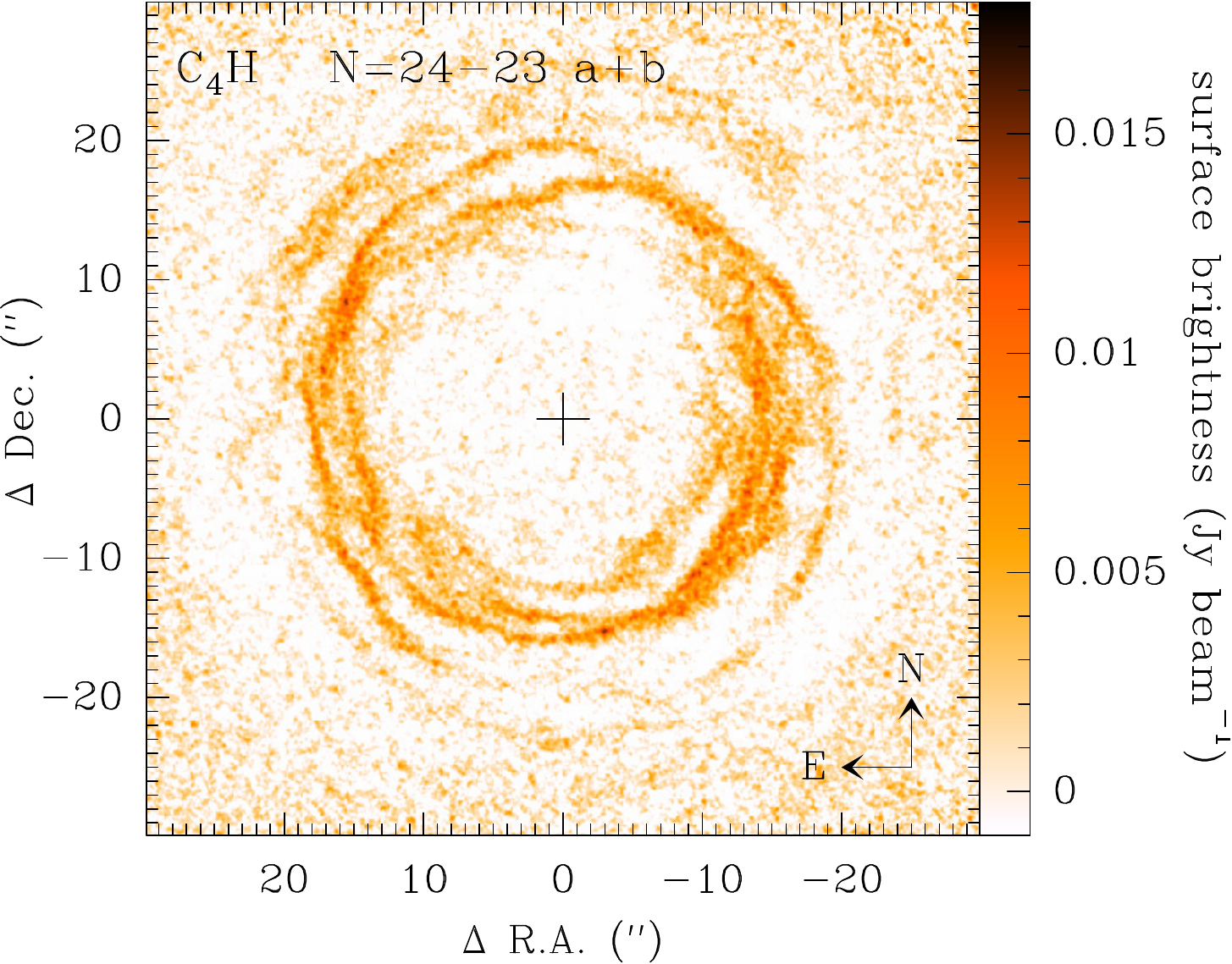}
\caption{ALMA Cycle\,2 brightness distribution of the $N=24-23$ transition of C$_4$H at $V_{\rm LSR}=V_{\rm sys}$, where the two spin-rotation components have been stacked. The synthesized beam is $0\secp37\times0\secp34$ and the rms per 0.62 km s$^{-1}$ channel is 4.8 mJy beam$^{-1}$. The map is centered on the position of the star, indicated by a cross.} \label{fig:map_c4h_24-23}
\end{figure}

The radicals C$_4$H and C$_3$N have several rotational transitions $N - (N-1)$ in the $\lambda~3$~mm band, each split into two closely spaced spin-rotation doublets of nearly equal intensity labeled as "a" and "b" (see Figs.~\ref{fig:map_c4h} and \ref{fig:map_c3n}), while the cyanopolyynes HC$_3$N and HC$_5$N also have several rotational transitions within the frequency range covered (see Fig.~\ref{fig:map_hc3n_hc5n}). Here we present the emission maps of each individual line and discuss whether the different lines of a given species have brightness distributions significantly different among them and whether the individual maps show any anomalies.

In Figs.~\ref{fig:map_c4h} and \ref{fig:map_c3n} we show the brightness distributions of each component of the doublet rotational transitions for the radicals C$_4$H ($N=9-8$ through $N=12-11$) and C$_3$N (from $N=9-8$ to $N=11-10$). The upper level energies of these transitions, in the range 20.5-35.6 K for C$_4$H and 21.4-31.3 K for C$_3$N, do not differ enough so to expect important differences in the excitation, which could in turn lead to significantly distinct emission distributions. Apart from some slight differences in the absolute intensities, the brightness distributions of the lines of C$_3$N are very similar both in the locations of the maxima and the arrangement of the arcs that shape the overall ring. The same is also true for C$_4$H, with the exception of the anomaly in the map of the $N=12-11$ $b$ transition. This line shows the same ring structure seen in the rest of the C$_4$H lines, but there is also a compact component centered on the star that is not resolved by the $1\secp14\times0\secp98$ synthesized beam. If the compact emission were to arise from C$_4$H, it would be very difficult to explain why it is only seen in this line while there is no hint of it in the maps of the other seven lines of C$_4$H. Instead this compact emission arises from CO $J=1-0$ in its first vibrationally excited state, whose rest frequency, 115221.754 MHz, is only 0.7 MHz away from that of the $N=12-11$ $b$ transition of C$_4$H. Emission from CO in the vibrational state $v=1$ had been detected in IRC\,+10216 in the pure rotational transitions $J=2-1$ and $J=3-2$ with the Submillimeter Array (SMA) by \cite{pat2009} (2009). The emission in these lines remains unresolved with an angular resolution of $\sim$2$''$, while that from the $J=1-0$ line we have detected remains unresolved with an angular resolution of $\sim$1$''$, indicating that the emission from CO $v=1$ arises from a region closer than $\sim$$10^{15}$~cm ($\sim$25~$R_*$) to the star. It is also worth noting that recently, emission from the pure rotational transition $J=3-2$ of CO in the vibrational state $v=1$ has been observed using ALMA with sub-arcsecond resolution in five additional AGB stars (\cite{kho2016} 2016), pointing to a line-formation region smaller than $\sim$10~$R_*$ in radius.

In Fig.~\ref{fig:map_c4h_24-23} we show the brightness distribution of C$_4$H in the rotational transition $N=24-23$, lying at 228.3 GHz. The map corresponds to an average of the two spin-rotation components in the central velocity channel and was obtained with ALMA band 6 during Cycle\,2. More details on these observations will be given elsewhere (Gu\'elin et al., in preparation). The map of this line shows the same structure of thin shells and inter-crossing arcs displayed by the $\lambda$~3 mm lines of C$_4$H, but seen at a higher angular resolution, $\sim$0.3$''$, and with a higher contrast because part of the extended emission filtered out by the interferometer has not been recovered owing to the lack of short-spacings data.

\begin{figure}[t]
\centering
\includegraphics[angle=0,width=\columnwidth]{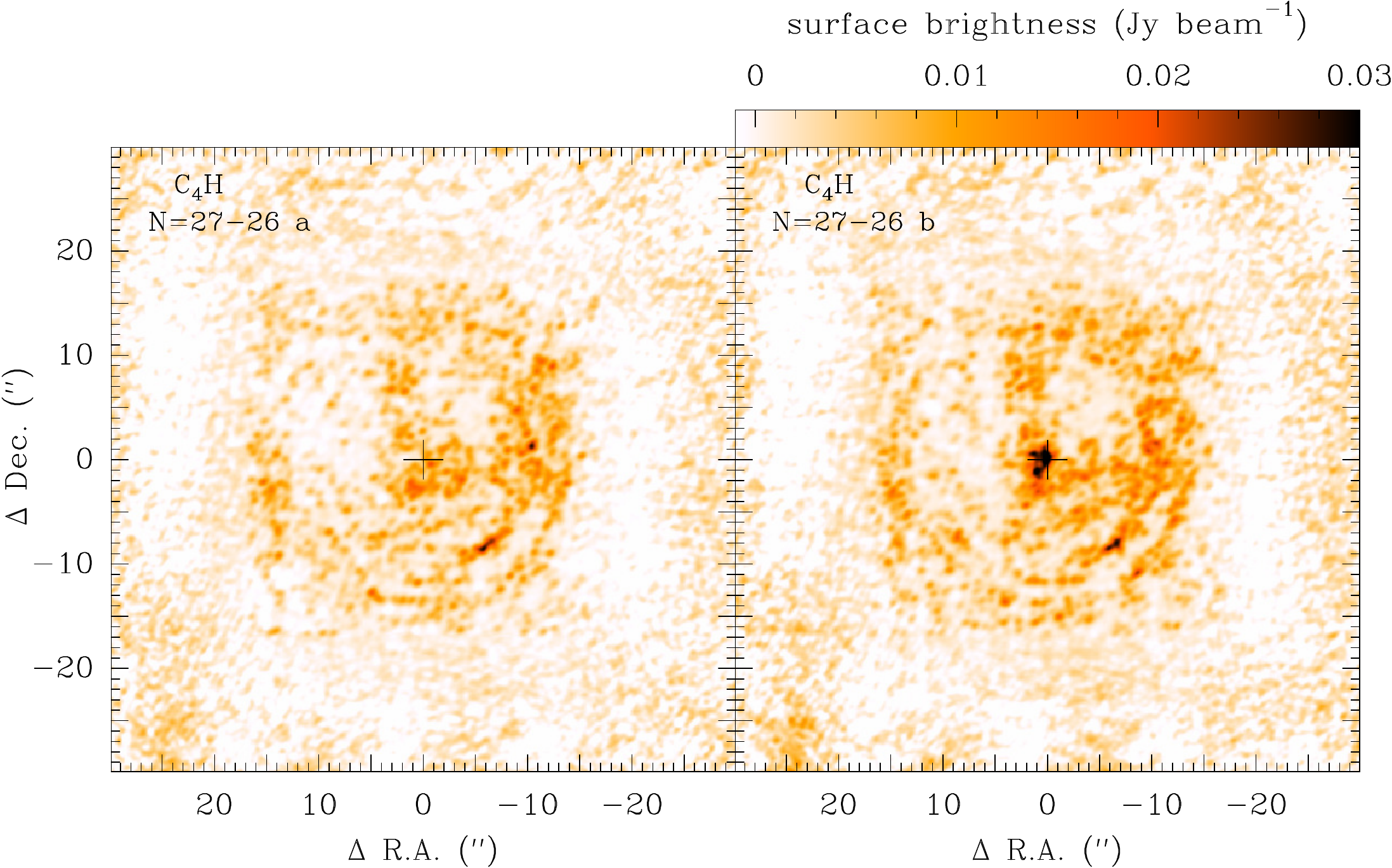}
\caption{ALMA Cycle\,0 brightness distributions of the two spin-rotation components of the $N=27-26$ transition of C$_4$H, averaged over the central 4 km s$^{-1}$ of each line. The synthesized beam is $0\secp77\times0\secp61$ and the rms per 4 km s$^{-1}$ channel is 1.3 mJy beam$^{-1}$. Maps are centered on the position of the star, indicated by a cross. Note that most of the ring structure is lost because this emission lies well outside the primary beam of $\sim$23$''$, and that a compact structure centered on the star is only clearly visible in the $N$ = 27-26 b line.} \label{fig:map_c4h_cycle0}
\end{figure}

\begin{figure*}
\centering
\includegraphics[angle=0,width=0.777\textwidth]{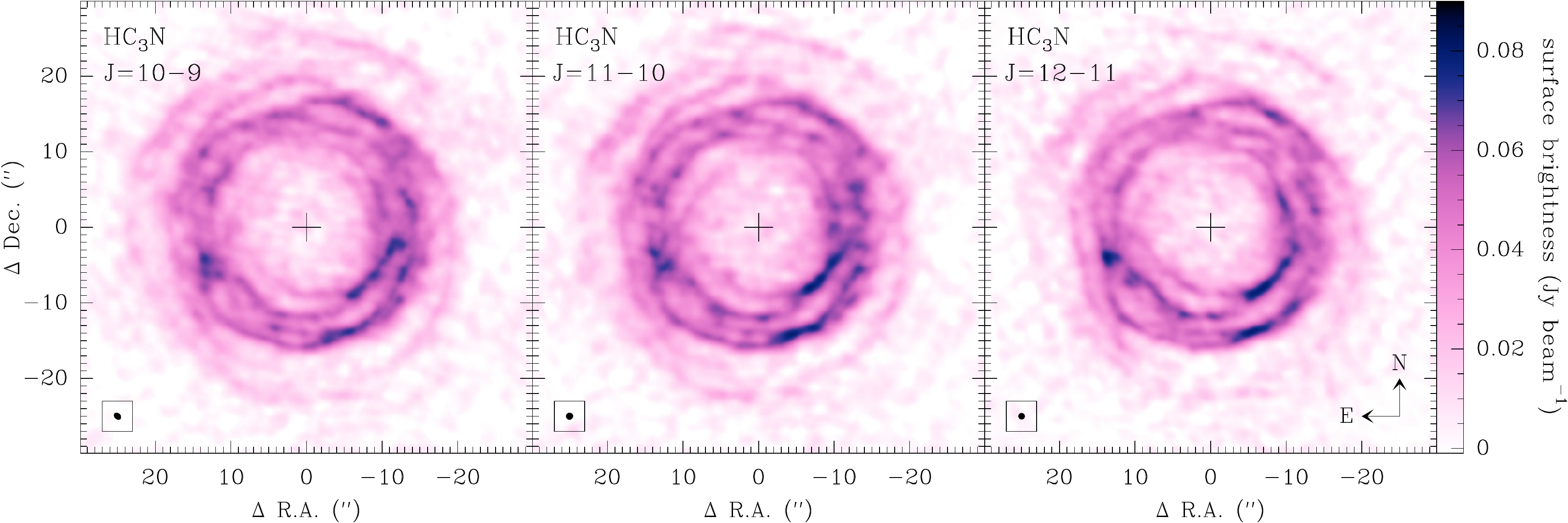} \includegraphics[angle=0,width=0.777\textwidth]{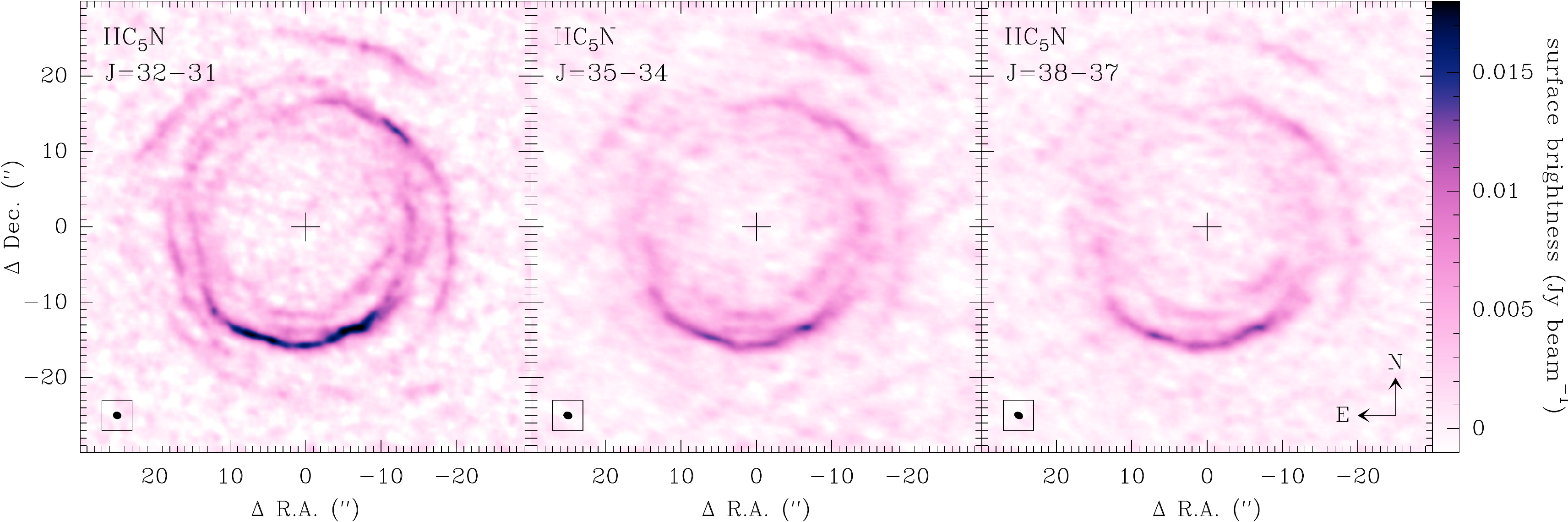}
\caption{Brightness distributions of the $J=10-9$, $J=11-10$, and $J=12-11$ lines of HC$_3$N (top panels), and $J=32-31$, $J=35-34$, and $J=38-37$ lines of HC$_5$N (bottom panels), averaged over the central 3 km s$^{-1}$ of each line. The size ($\sim1''$ for all lines) and shape of the synthesized beam is shown in the bottom left corner of each panel. The maps are centered on the position of the star, indicated by a cross.} \label{fig:map_hc3n_hc5n}
\end{figure*}

According to the ALMA maps of the four doublet lines of C$_4$H shown in Fig.~\ref{fig:map_c4h} and the transition shown in Fig.~\ref{fig:map_c4h_24-23}, the emission of C$_4$H is clearly restricted to the outer hollow shell, with no emission arising from the surroundings of the star. \cite{coo2015} (2015), however, find in their SMA maps of the doublet line $N=27-26$ of C$_4$H that there is compact emission centered on the star. This doublet, lying at 256.9 GHz, has been covered in our ALMA Cycle\,0 observations of IRC\,+10216, a large portion of which has been already published (\cite{cer2013} 2013; \cite{vel2015} 2015; \cite{agu2015} 2015; \cite{qui2016} 2016). As shown in Fig.~\ref{fig:map_c4h_cycle0}, at velocities around $V_{\rm LSR}=V_{\rm sys}$ the outer ring is barely seen in our ALMA Cycle\,0 data because the size of the ring is larger than the field of view ($\sim$23$''$) and thus this emission is not properly sampled. In contrast, compact emission around the star is clearly seen, but only in the $b$ component of the $N=27-26$ doublet and not in the $a$ component. This was not recognized in the data shown by \cite{coo2015} (2015) because these authors stacked the maps of the two components of the doublet to improve the sensitivity. As shown for the C$_4$H $N=12-11$ doublet in Fig.~\ref{fig:map_c4h}, it is more likely that the compact emission seen in the map of the $N=27-26$ $b$ component does not arise from C$_4$H but rather from a different carrier that has a transition which accidentally coincides with the $N=27-26$ $b$ component of C$_4$H. At this stage we do not have a fully convincing assignment for this line, but the emission maps of the entire set of $\lambda$~3 and 1 mm lines of C$_4$H strongly support that this species is only present in the outer shell and not in regions close to the star.

There are several rotational transitions of cyanoacetylene and cyanodiacetylene in the $\lambda$~3~ mm band observed here. For HC$_3$N just three lines are covered, from $J=10-9$ to $J=12-11$. These transitions have similar upper level energies, in the range 24.0-34.1 K, and thus, like the carbon chain radicals, we do not expect great differences in their excitation, and thus in their emission distribution. The maps of brightness of these three lines, centered on the systemic velocity, are shown in the top panels of Fig.~\ref{fig:map_hc3n_hc5n}, where it can be seen that there are indeed no substantive differences.

For HC$_5$N, 12 rotational lines, ranging from $J=32-31$ through $J=43-42$, have been covered. These transitions involve upper levels with energies from 67.5 to 120.9 K, i.e., substantially higher than those involved in the HC$_3$N transitions. The emission maps of each individual line of HC$_5$N (see a sample in the lower panels of Fig.~\ref{fig:map_hc3n_hc5n}) show that there is a noticeable decline in intensity with increasing $J$. This decrease in line intensity can be understood as a consequence of a less efficient excitation of levels of increasing energy in the outer shell at $\sim$15$''$ from the star, where the gas kinetic temperature and the volume density of particles are expected to be just 10-30 K and $(1-5)\times10^4$ cm$^{-3}$. As such, one might reasonably expect some radial shift in the emission peak between low-$J$ and high-$J$ lines, the latter being more easily excited at shorter radii because of the higher temperatures and densities. At current angular resolution and sensitivity of the measurements, there is no evidence of such shift among the $\lambda$~3~mm lines of HC$_5$N, whose emission distributions appear quite similar. It therefore seems that the structure of density enhancements in the form of arcs and shells is much more important than differential excitation effects in shaping the brightness distributions. The same is found for C$_4$H, C$_3$N, and HC$_3$N, where the transitions that we mapped involve levels with more similar energies. Therefore we conclude that regardless of the particular $\lambda$~3~mm transition adopted for any of the carbon chains studied here, the brightness distribution is a good proxy of the spatial distribution of the species. \cite{kel2015} (2015) have recently presented preliminary results on $\sim$1$''$ angular resolution VLA maps of cm-wavelength rotational lines of carbon chains in IRC\,+10216. These transitions involve rotational levels with energies significantly below those associated to the $\lambda$~3 mm lines presented here, and thus it will be interesting to see if there are significant differences in the distribution of the emission of cm- and 3~mm-wavelength lines that could arise from excitation effects.

\end{appendix}


\begin{thebibliography}{}

\bibitem[Ag\'undez \& Cernicharo]{agu2006} Ag\'undez, M. \& Cernicharo, J. 2006, \apj, 650, 374
\bibitem[Ag\'undez et al.]{agu2008} Ag\'undez, M., Fonfr\'ia, J. P., Cernicharo, J., et al. 2008, \aap, 479, 493
\bibitem[Ag\'undez]{agu2009} Ag\'undez, M. 2009, PhD Thesis, Universidad Aut\'onoma de Madrid
\bibitem[Ag\'undez et al.]{agu2010} Ag\'undez, M., Cernicharo, J., \& Gu\'elin, M. 2010, \apj, 724, L133
\bibitem[Ag\'undez et al.]{agu2012} Ag\'undez, M., Fonfr\'ia, J. P., Cernicharo, J., et al. 2012, \aap, 543, A48
\bibitem[Ag\'undez et al.]{agu2015} Ag\'undez, M., Cernicharo, J., Quintana-Lacaci, G., et al. 2015, \apj, 814, 143
\bibitem[Alexander et al.]{ale1976} Alexander, A. J., Kroto, H. W., \& Walton, D. R. M. 1976, \jms, 62, 175
\bibitem[Audinos et al.]{aud1994} Audinos, P., Kahane, C., \& Lucas, R. 1994, \aap, 287, L5
\bibitem[Berteloite et al.]{ber2010} Berteloite, C., Le Picard, S. D., Balucani, N., et al. 2010, \pccp, 12, 3677
\bibitem[Bieging \& Tafalla]{bie1993} Bieging, J. H. \& Tafalla, M. 1993, \aj, 105, 576
\bibitem[Bizzocchi et al.]{biz2004} Bizzocchi, L., Degli Esposti, C., \& Botschwina, P. 2004, \jms, 225, 145
\bibitem[Bohlin et al.]{boh1978} Bohlin, R. C., Savage, B. D., \& Drake, J. F. 1978, \apjs, 224, 132
\bibitem[Brooke et al.]{bro2014} Brooke, J. S. A., Ram, R. S., Western, C. M., et al. 2014, \apjs, 210, 23
\bibitem[Broten et al.]{bro1978} Broten, N. W., Oka, T., Avery, L. W., et al. 1978, \apj, 223, L105
\bibitem[Brown \& Millar]{bro2003} Brown, J. M. \& Millar, T. J. 2003, \mnras, 339, 1041
\bibitem[Bujarrabal et al.]{buj1981} Bujarrabal, V., Gu\'elin, M., Morris, M., \& Thaddeus, P. 1981, \aap, 99, 239
\bibitem[Canosa et al.]{can1997} Canosa, A., Sims, I. R., Travers, D., et al. 1997, \aap, 323, 644
\bibitem[Canosa et al.]{can2007} Canosa, A., P\'aramo, A., Le Picard, S. D., \& Sims, I. R. 2007, \icarus, 187, 558
\bibitem[Cernicharo \& Gu\'elin]{cer1996} Cernicharo, J. \& Gu\'elin, M. 1996, \aap, 309, L27
\bibitem[Cernicharo et al.]{cer1999} Cernicharo, J., Yamamura, I., Gonz\'alez-Alfonso, E., et al. 1999, \apj, 526, L41
\bibitem[Cernicharo et al.]{cer2000} Cernicharo, J., Gu\'elin, M., \& Kahane, C. 2000, \aaps, 142, 181
\bibitem[Cernicharo]{cer2004} Cernicharo, J. 2004, \apj, 608, L41
\bibitem[Cernicharo et al.]{cer2013} Cernicharo, J., Daniel, F., Castro-Carrizo, A., et al. 2013, \apj, 778, L25
\bibitem[Cernicharo et al.]{cer2015} Cernicharo, J., Marcelino, N., Ag\'undez, M., \& Gu\'elin, M. 2015, \aap, 575, A91
\bibitem[Chastaing et al.]{cha1998} Chastaing, D., James, P. L., Sims, I. R., \& Smith, I. W. M. 1998, \fdis, 109, 165
\bibitem[Chastaing et al.]{cha2001} Chastaing, D., Le Picard, S. D., Sims, I. R., \& Smith, I. W. M. 2001, \aap, 365, 241
\bibitem[Cherchneff et al.]{che1993a} Cherchneff, I., Glassgold, A. E., \& Mamon, G. A. 1993, \apj, 410, 188
\bibitem[Cherchneff \& Glassgold]{che1993b} Cherchneff, I. \& Glassgold, A. E. 1993, \apj, 419, L41
\bibitem[Choi et al.]{cho2004} Choi, N., Blitz, M. A., McKee, K., et al. 2004, \cpl, 384, 68
\bibitem[Clarke \& Ferris]{cla1995} Clarke, D. W. \& Ferris, J. P. 1995, \icarus, 115, 119
\bibitem[Clary et al.]{cla2002} Clary, D. C., Buonomo, E., Sims, I. R., et al. 2002, \jpca, 106, 5541
\bibitem[Cooksy et al.]{coo2015} Cooksy, A. L., Gottlieb, C. A., Killiam, T. C., et al. 2015, \apjs, 216, 30
\bibitem[Cooper et al.]{coo1995} Cooper, G., Burton, G. R., \& Brion, C. E. 1995, \jesrp, 73, 139
\bibitem[Cordiner \& Millar]{cor2009} Cordiner, M. A. \& Millar, T. J. 2009, \apj, 697, 68
\bibitem[Daniel et al.]{dan2012} Daniel, F., Ag\'undez, M., Cernicharo, J., et al. 2012, \aap, 542, A37
\bibitem[Daugey et al.]{dau2008} Daugey, N., Caubet, P., Bergeat, A., et al. 2008, \pccp, 10, 729
\bibitem[Dayal \& Bieging]{day1995} Dayal, A. \& Bieging, J. H. 1995, \apj, 439, 996
\bibitem[De Beck et al.]{deb2012} De Beck, E., Lombaert, R., Ag\'undez, M., et al. 2012, \aap, 539, A108
\bibitem[Decin et al.]{dec2010} Decin, L., Ag\'undez, M., Barlow, M. J., et al. 2010, \nature, 467, 64
\bibitem[Decin et al.]{dec2015} Decin, L., Richards, A. M. S., Neufeld, D., et al. 2015, \aap, 574, A5
\bibitem[Deguchi \& Uyemura]{deg1984} Deguchi, S. \& Uyemura, M. 1984, \apj, 285, 153
\bibitem[DeLeon \& Muenter]{del1985} DeLeon, R. L. \& Muenter, J. S. 1985, \jcp, 82, 1702
\bibitem[Dinh-V-Trung \& Lim]{din2008} Dinh-V-Trung \& Lim, J. 2008, \apj, 678, 303
\bibitem[Draine]{dra1978} Draine, B. T. 1978, \apjs, 36, 595
\bibitem[Dumouchel et al.]{dum2010} Dumouchel, F., Faure, A., Lique, F. 2010, \mnras, 406, 2488
\bibitem[Eichelberger et al.]{eic2007} Eichelberger, B., Snow, T. P., Barckholtz, C., \& Bierbaum, V. M. 2007, \apj, 667, 1283
\bibitem[Ferradaz et al.]{fer2009} Ferradaz, T., B\'enilan, Y., Fray, N., et al. 2009, \psp, 57, 10
\bibitem[Fonfr\'ia et al.]{fon2008} Fonfr\'ia, J. P., Cernicharo, J., Ritcher, M. J., \& Lacy, J. H. 2008, \apj, 673, 445
\bibitem[Fournier]{fou2014} Fournier, M. 2014, PhD Thesis, Universit\'e de Rennes
\bibitem[Fray et al.]{fra2010} Fray, N., B\'enilan, Y., Gazeau, M.-C., et al. 2010, \jgr, 115, E06010
\bibitem[Fukuzawa \& Osamura]{fuk1997} Fukuzawa, K. \& Osamura, Y. 1997, \apj, 489, 113
\bibitem[Fukuzawa et al.]{fuk1998} Fukuzawa, K., Osamura, Y., \& Schaefer III, H. F. 1998, \apj, 505, 278
\bibitem[Glassgold et al.]{gla1986} Glassgold, A. E., Lucas, R., \& Omont, A. 1986, \aap, 157, 35
\bibitem[Gonz\'alez-Alfonso et al.]{gon2007} Gonz\'alez-Alfonso, E., Neufeld, D. A., \& Melnick, G. J. 2007, \apj, 669, 412
\bibitem[Gottlieb et al.]{got1983} Gottlieb, C. A., Gottlieb, E. W., Thaddeus, P., \& Kawamura, H. 1983, \apj, 275, 916
\bibitem[Green \& Chapman]{gre1978} Green, S. \& Chapman, S. 1978, \apjs, 37, 169
\bibitem[Groenewegen et al.]{gro2012} Groenewegen, M. A. T., Barlow, M. J., Blommaert, J. A. D. L., et al. 2012, \aap, 543, L8
\bibitem[Gu et al.]{gu2009} Gu, X., Kim, Y. S., Kaiser, R. I., et al. 2009, \pnas, 106, 16078
\bibitem[Gu\'elin et al.]{gue1978} Gu\'elin, M., Green, S., \& Thaddeus, P. 1978, \apj, 224, L27
\bibitem[Gu\'elin et al.]{gue1993} Gu\'elin, M., Lucas, R., \& Cernicharo, J. 1993, \aap, 280, L19
\bibitem[Gu\'elin et al.]{gue1997} Gu\'elin, M., Lucas, R., \& Neri, R. 1997, in CO: Twenty-Five Years of Millimeter Wave Spectroscopy, ed. W. B. Latter et al. (Dordrecht: Kluwer), IAU Symp., 170, 359
\bibitem[Gu\'elin et al.]{gue1999} Gu\'elin, M., Neininger, N., Lucas, R., \& Cernicharo, J. 1999, in The Physics and Chemistry of the Interstellar Medium, Proc. 3rd Cologne-Zermatt Symp., ed. V. Ossenkopf, et al. (Herdecke: GCA-Verlag), 326
\bibitem[Gupta et al.]{gup2009} Gupta, H., Gottlieb, C. A., McCarthy, M. C., \& Thaddeus, P. 2009, \apj, 691, 1494
\bibitem[Halpern et al.]{hal1988} Halpern, J. B., Miller, G. E., \& Okabe, H. 1988, \jppa, 42, 63
\bibitem[Heays et al.]{hea2017} Heays, A. N., Bosman, A. D., \& van Dishoeck, E. F. 2017, \aap, in press
\bibitem[Herbst \& Leung]{her1989} Herbst, E. \& Leung, C. M. 1989, \apjs, 69, 271
\bibitem[Hoobler \& Leone]{hoo1997} Hoobler, R. J. \& Leone, S. R. 1997, \jgr, 102, 28717
\bibitem[Huang et al.]{hua2000} Huang, L. C. L., Asvany, O., Chang, A. H. H., et al. 2000, \jcp, 113, 8656
\bibitem[H\"ubner et al.]{hub2005} H\"ubner, M., Castillo, M., Davies, P. B., \& R\"pcke, J. 2005, \scaa, 61, 57
\bibitem[Hudson]{hud1971} Hudson, R. D. 1971, \rgsp, 9, 305
\bibitem[Jolly et al.]{jol2007} Jolly, A., B\'enilan, Y., \& Fayt, A. 2007, \jms, 242, 46
\bibitem[Keller et al.]{kel2015} Keller, D., Menten, K. M., Kami\'nski, T., \& Claussen, M. J. 2015, in Why Galaxies Care about AGB Stars III, ed. F. Kerschbaum, et al., ASP Conf. Ser, 497, 123
\bibitem[Khouri et al.]{kho2016} Khouri, T., Vlemmings, W. H. T., Ramstedt, S., et al. 2016, \mnras, 463, L74
\bibitem[Killian et al.]{kil2007} Killian, T. C., Gottlieb, C. A., \& Thaddeus, P. 2007, \jcp, 127, 114320
\bibitem[Klisch et al.]{kli1995} Klisch, E., Klaus, Th., Belov, S. P., et al. 1995, \aap, 304, L5
\bibitem[Kloster-Jensen et al.]{klo1974} Kloster-Jensen, E., Haink, H.-J., \& Christen, H. 1974, \hca, 57, 1731
\bibitem[Kroto et al.]{kro1987} Kroto, H. W., Heath, J. R., Obrien, S. C., et al. 1987, \apj, 314, 352
\bibitem[Landera et al.]{lan2008} Landera, A., Krishtal, S. P., Kislov, V. V., et al. 2008, \jcp, 128, 214301
\bibitem[Le Petit et al.]{lep2006} Le Petit, F. L., Nehm\'e, C., Le Bourlot, J., \& Roueff, E. 2006, \apjs, 164, 506
\bibitem[Leach et al.]{lea2014} Leach, S. Garc\'ia, G. A., Mahjoub, A., et al. 2014, \jcp, 140, 174305
\bibitem[Le\~ao et al.]{lea2006} Le\~ao, I. C., de Laverny, P., M\'ekarnia, D., et al. 2006, \aap, 455, 187
\bibitem[Li et al.]{li2014} Li, X., Millar, T. J., Walsh, C., et al. 2014, \aap, 568, A111
\bibitem[Linnartz et al.]{lin1999} Linnartz, H., Motylewski, T., Vaizert, O., et al. 1999, \jms, 197, 1
\bibitem[Lucas et al.]{luc1995} Lucas, R., Gu\'elin, M., Kahane, C., et al. 1995, \apss, 224, 293
\bibitem[Loison \& Bergeat]{loi2009} Loison, J.-C. \& Bergeat, A. 2009, \pccp, 11, 655
\bibitem[Loomis et al.]{loo2016} Loomis, R. A., Shingledecker, C. N., Langston, G., et al. 2016, \mnras, 463, 4175
\bibitem[Mauron \& Huggins]{mau1999} Mauron, N. \& Huggins, P. J. 1999, \aap, 349, 203
\bibitem[McCarthy et al.]{mcc1995} McCarthy, M. C., Gottlieb, C. A., Thaddeus, P., et al. 1995, \jcp, 103, 7820
\bibitem[McElroy et al.]{mce2013} McElroy, D., Walsh, C., Markwick, A. J., et al. 2013, \aap, 550, A36
\bibitem[Millar \& Herbst]{mil1994} Millar, T. J. \& Herbst, E. 1994, \aap, 288, 561
\bibitem[Millar et al.]{mil2000} Millar, T. J., Herbst, E., \& Bettens, R. P. A. 2000, \mnras, 316, 195
\bibitem[Morris et al.]{mor1976} Morris, M., Turner, B. E., Palmer, P., \& Zuckerman, B. 1976, \apj, 205, 82
\bibitem[M\"uller et al.]{mul2000} M\"uller, H. S. P., Klaus, T., \& Winnewisser, G. 2000, \aap, 357, L65
\bibitem[Nejad \& Millar]{nej1987} Nejad, L. A. M. \& Millar, T. J. 1987, \aap, 183, 279
\bibitem[Nuth \& Glicker]{nut1982} Nuth, J. A. \& Glicker, S. 1982, \jqsrt, 28, 223
\bibitem[P\'aramo et al.]{par2008} P\'aramo, A., Canosa, A., Le Picard, S. D., \& Sims, I. R. 2008, \jpca, 112, 9591
\bibitem[Pardo et al.]{par2005} Pardo, J. R., Cernicharo, J., \& Goicoechea, J. R. 2005, \apj, 628, 275
\bibitem[Patel et al.]{pat2009} Patel, N. A., Young, K. H., Br\"unken, S., et al. 2009, \apj, 691, L55
\bibitem[Patel et al.]{pat2011} Patel, N. A., Young, K. H., Gottlieb, C. A., et al. 2011, \apjs, 193, 17
\bibitem[Pety et al.]{pet2005} Pety, J., Teyssier, D., Foss\'e, D., et al. 2005, \aap, 435, 885
\bibitem[Pulliam et al.]{pul2011} Pulliam, R. L., Edwards, J. L., \& Ziurys, L. M. 2011, \apj, 743, 36
\bibitem[Quintana-Lacaci et al.]{qui2016} Quintana-Lacaci, G., Cernicharo, J., Ag\'undez, M., et al. 2016, \apj, 818, 192
\bibitem[Schwell et al.]{sch2012} Schwell, M., B\'enilan, Y., Fray, N., et al. 2012, \molphys, 110, 21
\bibitem[Seki et al.]{sek1996a} Seki, K., Yagi, M., He, M., et al. 1996a, \cpl, 258, 657
\bibitem[Seki et al.]{sek1996b} Seki, K., He, M., Liu, R., \& Okabe, H. 1996b, \jpc, 100, 5349
\bibitem[Shindo et al.]{shi2003} Shindo, F., B\'enilan, Y., Guillemin, J.-C., et al. 2003, \pss, 51, 9
\bibitem[Silva et al.]{sil2008} Silva, R., Gichuhi, W. K., Huang, C., et al. 2008, \pnas, 105, 12713
\bibitem[Silva et al.]{sil2009} Silva, R., Gichuhi, W. K., Kislov, V. V., et al. 2009, \jpca, 113, 11182
\bibitem[Sims et al.]{sim1993} Sims, I. R., Queffelec, J.-L., Travers, D., et al. 1993, \cpl, 211, 461
\bibitem[Sun et al.]{sun2015} Sun, Y.-L., Huang, W.-J., \& Lee, S.-H. 2015, \jpcl, 6, 4117
\bibitem[Tarroni \& Carter]{tar2004} Tarroni, R. \& Carter, S. 2004, \molphys, 102, 2167
\bibitem[Thompson \& Dalby]{tho1968} Thompson, R. \& Dalby, F. W. 1968, \cjp, 46, 53
\bibitem[Thorwirth et al.]{tho2000} Thorwirth, S., M\"uller, H. S. P., \& Winnewisser, G. 2000, \jms, 204, 133
\bibitem[Uyemura et al.]{uye1982} Uyemura, M., Deguchi, S., Nakada, Y., \& Onaka, T. 1982, \bcsj, 55, 384
\bibitem[Velilla Prieto et al.]{vel2015} Velilla Prieto, L., Cernicharo, J., Quintana-Lacaci, G., et al. 2015, \apj, 805, L13
\bibitem[Wakelam et al.]{wak2015} Wakelam, V., Loison, J.-C., Herbst, E., et al. 2015, \apjs, 217, 20
\bibitem[Walsh et al.]{wal2009} Walsh, C., Harada, N., Herbst, E., \& Millar, T. J. 2009, \apj, 700, 752
\bibitem[Winnewisser \& Walmsley]{win1978} Winnewisser, G. \& Walmsley, C. M. 1978, \aap, 70, L37
\bibitem[Woon]{woo1995} Woon, D. E. 1995, \cpl, 244, 45
\bibitem[Zhang et al.]{zha2009} Zhang, F., Seol, K., Kaiser, R. I., et al. 2009, \jcp, 130, 234308

\end{thebibliography}
\end{document}